\documentclass[11pt,preprint]{aastex}
\usepackage{amsmath, amssymb, graphicx,times,rotating}
\newcommand\numberthis{\addtocounter{equation}{1}\tag{\theequation}}

\usepackage{natbib}
\newcommand{\disp}{\displaystyle}
\begin{document}

%\shorttitle{Jet variability models}
%\title{Models for variability of relativistic jets from AGN}
\title{Kinematics of and emission from helically orbiting blobs in a relativistic magnetized jet}

\author{P. Mohan\altaffilmark{1,2 \ \dagger} \& A.\ Mangalam\altaffilmark{1, \ddagger}}

\altaffiltext{1}{Indian Institute of Astrophysics, Sarjapur Road, Koramangala, Bangalore, 560034, India}

\altaffiltext{2}{Aryabhatta Research Institute of Observational Sciences, Manora Peak, Nainital, 263002, India}

\email{prashanth@iiap.res.in$^\dagger$, mangalam@iiap.res.in$^\ddagger$}

\begin{abstract}
We present a general relativistic (GR) model of jet variability in active galactic nuclei due to orbiting blobs in helical motion along a funnel or cone shaped magnetic surface anchored to the accretion disk near the black hole. Considering a radiation pressure driven flow in the inner region, we find that it stabilizes the flow, yielding Lorentz factors ranging between 1.1 and 7 at small radii for reasonable initial conditions. Assuming these as inputs, simulated light curves (LCs) for the funnel model include Doppler and gravitational shifts, aberration, light bending, and time delay. These LCs are studied for quasi-periodic oscillations (QPOs) and the power spectral density (PSD) shape and yield an increased amplitude ($\sim$ 12 \%); a beamed portion and a systematic phase shift with respect to that from a previous special relativistic model. The results strongly justify implementing a realistic magnetic surface geometry in Schwarzschild geometry to describe effects on emission from orbital features in the jet close to the horizon radius. A power law shaped PSD with a typical slope of $-2$ and QPOs with timescales in the range of $(1.37 - 130.7)$ days consistent with optical variability in Blazars, emerges from the simulations for black hole masses $M_{\bullet} = (0.5 - 5) \times 10^8 M_{\odot}$ and initial Lorentz factors $\gamma_{jet,i} = 2 - 10$. The models presented here can be applied to explain radio, optical, and X-ray variability from a range of jetted sources including active galactic nuclei, X-ray binaries and neutron stars.

\end{abstract}


\section{Introduction}

Bulk flowing plasma close to the innermost stable circular orbit (ISCO) of a black hole consisting of the accretion disk and developing jet is strongly influenced by magnetic field structures, differential rotation of the disk, and radiative structuring. The flow velocity field and magnetic fields are dominated by the azimuthal and possibly the radial components. The poloidal components are non-zero owing to vertical structuring, random motions, radiation pressure, and dynamo action in the disk. The magnetic field strength generated by dynamo action in the inner disk is expected to be in equipartition with the gas pressure (e.g. \citealt{1995ApJ...446..741B}), allowing for a reasonably well developed vertically structuring in the field \citep{1996astro.ph..2022S}.

Variability is often observed in emission from active galactic nuclei (AGNs) with strong jet components such as blazars, inferred from the domination of synchrotron processes in radio to optical light curves (LCs) and spectra from regions at parsec scales as well as close to the base of the jet (e.g. \citealt{1996AJ....111.2187W,2012MNRAS.425.1357G}). Theoretical models are often applicable to emission from regions which are some distance away from the central region where structures are resolvable. Some of these include Doppler beaming of a stream or blobs of plasma accelerated to relativistic velocities along helical paths \citep{1992A&A...255...59C,1995A&A...302..335S,2004ApJ...615L...5R} as was recently observed in the blazar BL Lacertae \citep{2008Natur.452..966M}, and shocks propagating along the relativistic jet (\citealt{1985ApJ...298..114M}) which explains the variability in some blazars (e.g. \citealt{1991A&A...241...15Q,2008ApJ...672...40H}). Quasi-periodic variability can be caused due to the orbital motion of the flow along helical trajectories which get beamed when their local angle is close to the angle to the observer's line of sight in the inner region where the jet is just developing. This effect is expected to last for a few cycles.

In a study of X-ray binaries, the inner jet appears to be intrinsically linked to the corona as spectral characteristics of its emissions are the same as those from a Comptonized corona \citep{2005ApJ...635.1203M}. In another study, VLBI observations at 1.3 mm resolve structures and study the emission from the inner jet of M87 \citep{2012Sci...338..355D}. In the study, a direct scaling relation between the angular size and the distance to the object indicates that the region of emission is very compact, even inside of the ISCO (5.5 $\pm$ 0.4 $R_{\mathrm{s}}$,  where $R_{\mathrm{s}}$ is the Schwarzschild radius), implying that the emitting source is on prograde orbits. These and similar studies indicate that there is a strong disk-jet connection, which is also likely as perturbations produced in the disk can be advected into the jet and amplified there via Doppler boosting (e.g. \citealt{2006ASPC..350..183W}).

Emission in AGNs such as blazars is observed in a wide variety of wavelengths ranging from radio to optical to Gamma-rays (e.g. \citealt{2011ApJ...726...43A}) and can be inferred in many cases to arise from the jet. Intra-day variability in the optical bands is frequently observed in strongly jetted radio loud AGNs. The measured magnitude often changes by an order or more in less than a day (e.g. \citealt{2008AJ....135.1384G}). Jet variability in the X-rays is observed on timescales of a few hours. A possible 4.6 hr QPO in a X-ray light curve of the blazar PKS 2155-304 and the interpretation is discussed in terms of instabilities in the disk being advected into the jet and intrinsic jet based processes including shocks in jets \citep{2009A&A...506L..17L}. IDV timescales ranging between 15.7 and 46.8 ks have been detected in eight light curves from 1ES 1426+428 and PKS 2155-304 \citep{2010ApJ...718..279G}. Intra-night variability in the optical B and R bands have detected timescales ranging from several hours to a week in the radio loud narrow line Seyfert 1 galaxy SDSS J094857.3+002225, interpreted in terms of physical processes in a relativistic jet \citep{2010ApJ...715L.113L}. These and other studies indicate that short-term (few hours) to long-term (a few days to weeks) variability is commonly observed in the radio, optical, and X-ray wavelengths. A helical kinematical jet model was applied to explain the trajectories of blobs in the inner jet \citep{1995A&A...302..335S}. In the study, the model is applied to quasar 3C 345 to infer an inclination angle of $6.8^{\circ}$ and typical Lorentz factors of 5.8 and 4.6 for two radio components.

In our paper we define ``blobs'' to be density inhomogeneities which are acted upon by the centrifugal force, radiation pressure and drag, and the gravitational force of the black hole. We however do not include any thermodynamic evolution of the blob. The blob flows along with the bulk plasma in the region with a dominant magnetic field, where the plasma is constrained by flux freezing to flow along the magnetic surface encompassing the jet. The variability could be attributed to orbital blobs propagating along the jet, the emission from which is beamed along the observer line of sight causing rapid changes in the received flux. Our model assumes a flow which has transited from the accretion disk onto the jet through the magnetic field surface anchored onto the disk at foot points close to the black hole.  Thus, blobs such as orbiting spots could be present in the jet at various scales in the mass loaded jet. This is expected to occur regardless of the jet models such as the purely electro-magnetic jets or the magneto-hydrodynamic jets. The developed model is used to study the quasi-periodic oscillations (QPOs) phenomenon in addition to the timing study of simulated light curves using the Fourier periodogram and wavelet analysis. Our model can be used to study both the timing properties and the variable emission from orbiting blobs as well as model the jet structure and place constraints on the trajectories of these orbiting blobs which can be compared with observations.

Earlier works in this direction have mainly addressed only accretion disk based variability and its timing properties. Models include the variability due to orbiting blobs and are applicable to optical/UV and X-ray wavelengths (e.g., \citealt{1991A&A...246...21Z,1993ApJ...406..420M}). Recent models also include the effect of aberration based on the observer in a local static frame \citep{2005A&A...441..855P}, hinting at the necessity for a full general relativistic treatment in the disk-jet case. A model for the QPO and its timing properties such as the quality factor, the break frequency inferred from the PSD shape for a general relativistic thin disk was presented in (\citealt{2014ApJ...791...74M}). There is a necessity for a treatment of the emission source kinematics and light ray paths in curved space-time as the inner jet is close to the black hole; also general relativistic effects on this emission have not been treated in earlier works. A preliminary study was presented in (\citealt{2014PhDT...PM,2014JApA...35..431M}).

\begin{figure}[h!]
\centerline{\includegraphics[scale=0.51]{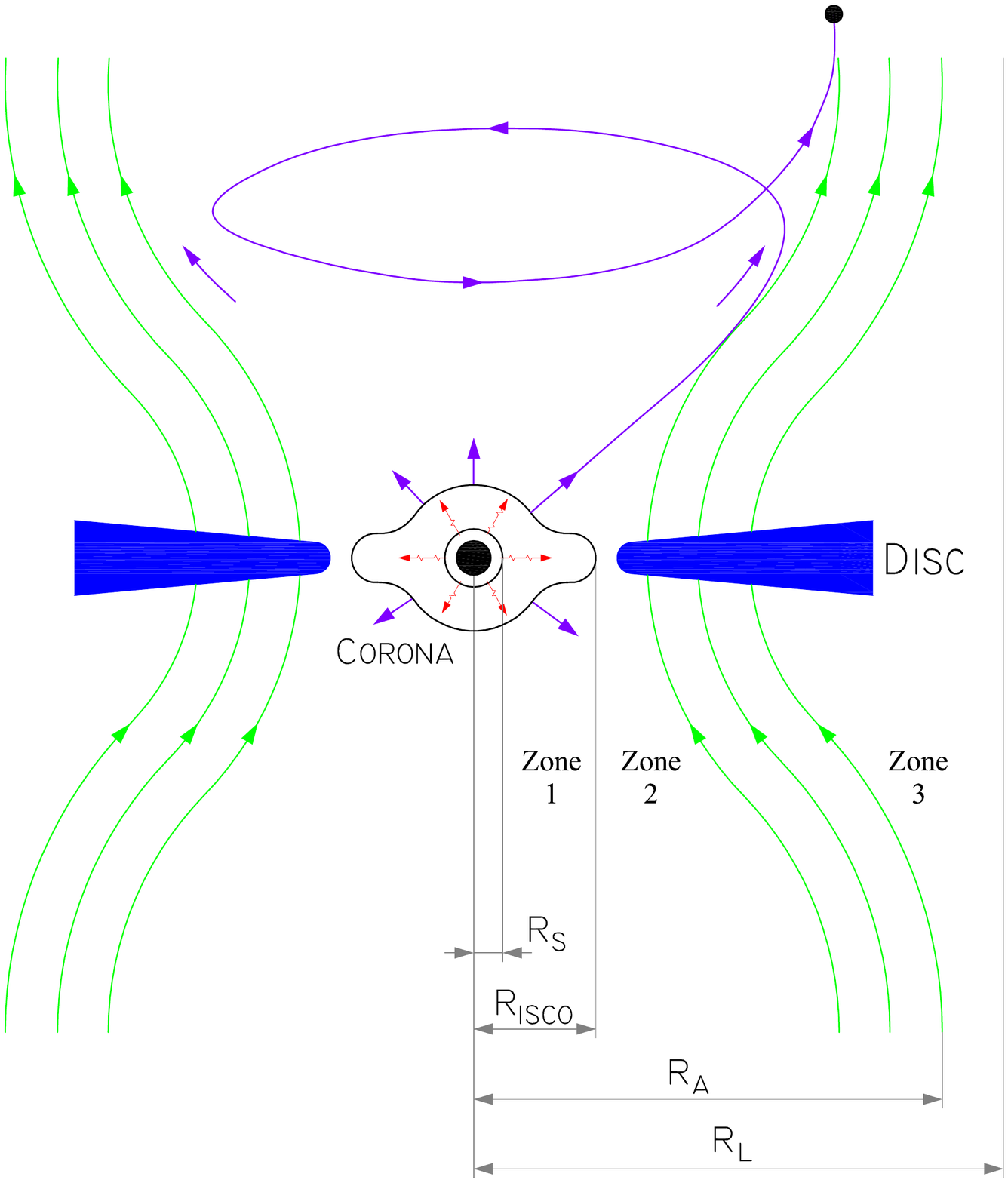}}
\caption{A schematic indicating the kinematically different zones in the trajectory of a typical blob. The black hole with Schwarzschild radius $R_s = 2 G M_\bullet/c^2$ is at the centre. In the region ($R_s < r < R_{ISCO}$) marked Zone 1, radiation pressure is dominant and drives the outflowing blobs sourced from a zero angular momentum corona. In the transition region ($R_{ISCO} < r < R_A$) marked Zone 2, the kinematics of the flow are driven by radiation as well as the co-rotating magnetic field lines that are anchored to the disk. In Zone 3, at radii $R_A < r < R_L$, where $R_A$ is the Alfv\'{e}n radius and $R_L$ is the radius of the light cylinder, the inertia dominates and the orbital angular momentum of the blob has reached an asymptotic value. {\em See the online article for the color version.}}
\label{RadiationGeometry}
\end{figure}

A schematic indicating the basic features of the wind in the context of the black hole system is shown in Fig. \ref{RadiationGeometry}. The geometry consists of three zones. In  the region ($R_s < r < R_{ISCO}$) marked as Zone 1, the radiation pressure is dominant and drives the outflowing blobs which are sourced from a hot corona (thermal temperature of $\sim$ 85 keV, e.g. \citealt{2014ApJ...791...74M}) with zero angular momentum (as the inflow towards the black hole would be on plunging orbits). We consider the effects of radiation, namely pressure and drag in a two dimensional model in \S \ref{radpr}, thereby extending the one-dimensional model of \cite{1990ApJ...361..470A}. In \S \ref{tframework}, we construct a model in Schwarzschild geometry for the  kinematics and emission of the blob for the region ($R_A < r < R_L$) marked as Zone 3 and beyond ($r > R_L$). The transition region ($R_{ISCO} < r < R_A$) marked as Zone 2 requires a treatment of the relativistic Grad-Shafranov to self-consistently describe the geometry of the field and kinematics of the flow. While we postpone this for future studies, in this paper we take the results from \S \ref{radpr} to provide the launch parameters for calculating the blob trajectory in a conical or funnel geometry for the magnetized jet in \S \ref{kinmodels}. We assume that the blob which is accelerated by radiation reaches the magnetic surface where it is centrifugally driven to higher Lorentz factors by the co-rotation with the foot point. In the analysis in \S \ref{kinmodels} applicable to Zone 3, we calculate typical Lorentz factors which can be obtained both from the initial acceleration due to radiation in Zone 1 and co-rotation in Zone 2; in Zone 3 we assume that the blob has reached a final angular momentum achieved at the Alfv\'{e}n point; $j_\infty = \varpi^2_A \Omega$ where $\varpi_A$ is the cylindrical radius of the Alfv\'{e}n point and $\Omega$ is the spin of the magnetic surface. Based on the kinematics of the model, the  emission is also calculated for the centrifugally driven flow on an assumed conical or funnel shaped geometry for the magnetic surface. While this is a simplified approach which improves upon previous work, we plan to construct a fully self consistent model for Zone 2 in the future. We calculate the trajectory of the orbiting blob and the expected light curve from a special relativistic cone model as presented in \cite{1992A&A...255...59C} and then from a general relativistic cone model and a realistic funnel model given the instantaneous position and velocity in each case. Results from the analysis of simulated LCs include addressing the QPO phenomenon, its evolution and the shape of the power spectral density for multiple emitting regions are presented in \S \ref{results}. We then discuss the advantages of our general relativistic funnel model and interpret the results of the simulations in \S \ref{conclusions}.

\section{Radiation driven wind in the launching region}
\label{radpr}

The effect of radiation pressure on radially outflowing relativistic particles have been studied by \cite{1990ApJ...361..470A}. While \cite{1991A&A...252..835V} included the black hole spin on the radiation pressure driven radial motion of particles,  the effects of  thermodynamic evolution was included by \cite{2006PASJ...58..203H}.

In addition to the effects of radiation pressure considered in (\citealt{1990ApJ...361..470A}), we extend their one dimensional model to the general case of three dimensions (that  simplifies to two for the resulting planar trajectories in a spherically symmetric metric) by including effects of radiation drag arising from the azimuthal and the $\theta$ components of the equations of motion. The four acceleration of the blobs in Schwarzschild geometry is set equal to the radiation forces of pressure and drag on the blob and hence the Lorentz factors are derived. 

The observer geometry and the vectors is presented in Fig. 3 of (\citealt{1990ApJ...361..470A}). First, we treat the blob motion in full $(r,\theta,\phi)$ spherical coordinates. The evolution of the velocity due to the action of the radiation pressure and drag self consistently governs the trajectory of the blobs until it reaches the launch positions on the magnetic surface at larger distances. The region where this occurs is radiation dominated and consists of a zero angular momentum hot flow sourced from a corona. The kinematics of this region, Zone 1, is  of interest in the current section. 

The model is cast in a Schwarzschild space-time with the line element,

\begin{equation}
ds^2 = -(1-2 M/r) c^2 dt^2+\frac{dr^2}{(1-2 M/r)}+r^2 (d \theta^2+\sin^2\theta d \phi^2),
\label{metric}
\end{equation}
where $(r,\theta,\phi)$ are the spherical polar coordinates, $M = G M_{\bullet}/c^2$ is the gravitational radius with $M_{\bullet}$ being the black hole mass. The covariant components of this diagonal metric, expressed in these coordinates are given by
\begin{equation}
g_{\alpha \beta} = (g_{tt},g_{rr},g_{\theta \theta},g_{\phi \phi}) = \left(-(1-2 M/r),\frac{1}{1-2 M/r},r^2,r^2 \sin^2\theta \right).
\end{equation}

%The tetrads for Schwarzschild geometry are given by,
%\begin{align} \label{refframes}
%&e_{t}^{(t)} = (1-2 M/r)^{1/2}; ~ e_{r}^{(r)} = (1-2 M/r)^{-1/2}; ~ e_{\theta}^{(\theta)} = r; ~ e_{\phi}^{(\phi)} = r \sin \theta \\ \nonumber
%&e^{t}_{(t)} = (1-2 M/r)^{-1/2}; ~e^{r}_{(r)} = (1-2 M/r)^{1/2}; ~ e^{\theta}_{(\theta)} = r^{-1}; ~ e^{\phi}_{(\phi)} = (r \sin \theta)^{-1}
%\end{align}

%which are obtained by writing the metric in a tetrad notation. 

By writing the line element in the form 
\begin{align} \label{ds2}
ds^2 = -c^2 dt^2 (1-2 M/r) &\left(1-(1-2 M/r)^{-2} \left(\frac{dr}{dt}\right)^2-r^2 (1-2 M/r)^{-1} \left(\frac{d\theta}{dt}\right)^2 \right.\\ \nonumber
&\left.-r^2 \sin^2 \theta (1-2 M/r)^{-1} \left(\frac{d\phi}{dt}\right)^2\right),
\end{align}

we can write the blob three-velocity components as
\begin{align}\label{betadt}
\beta_r &= (1-2 M/r)^{-1} \frac{1}{c} \frac{dr}{dt} \\ \nonumber
\beta_\theta &= (1-2 M/r)^{-1/2} \frac{r}{c} \frac{d\theta}{dt} \\ \nonumber
\beta_\phi &= (1-2 M/r)^{-1/2} \frac{r \sin \theta}{c} \frac{d\phi}{dt} .
\end{align}

With the above representation, the proper time $d\tau^2 = -ds^2$ can be written as
\begin{equation}
d\tau^2 = c^2 dt^2 \gamma^{-2} (1-2 M/r),
\end{equation}
where $\gamma = (1-\beta^2_r-\beta^2_\theta-\beta^2_\phi)^{-1/2}$ is the Lorentz factor. The components of the four-velocity of the emitting blob along the jet $u^{\alpha} = dx^{\alpha}/d\tau = (dx^{\alpha}/dt) (dt/d\tau)$ are then
\begin{equation}
u^{\alpha} = \left(\frac{1}{c} (1-2M/r)^{-1/2} \gamma,(1-2M/r)^{1/2} \beta_r \gamma,\frac{1}{r} \beta_\theta \gamma,\frac{1}{r \sin \theta} \beta_\phi \gamma\right),
\label{4vel}
\end{equation}

The covariant components of the four-velocity can be obtained by lowering the indices using the metric $u_{\beta} = u^{\alpha} g_{\alpha\beta}$, 
\begin{equation}
u_{\beta} =  (-(1-2M/r)^{1/2} \gamma,(1-2M/r)^{-1/2} \beta_r \gamma, r \beta_\theta \gamma,r \sin \theta \beta_\phi \gamma).
\label{4velc}
\end{equation}

The contra-variant components of the acceleration are given by
\begin{equation}
a^\alpha = \frac{du^\alpha}{d\tau}+ \Gamma^{\alpha}_{\mu \nu} u^{\mu} u^{\nu}.
\end{equation}

The radial component is then given by
\begin{align}
a^r &= \frac{\partial u^r}{\partial r} u^r+\frac{\partial u^r}{\partial \theta} u^\theta+\frac{\partial u^r}{\partial \phi} u^\phi+\Gamma^{r}_{\mu \nu} u^{\mu} u^{\nu}\\ \nonumber
&= \frac{\partial u^r}{\partial r} u^r+\frac{\partial u^r}{\partial \theta} u^\theta+\frac{\partial u^r}{\partial \phi} u^\phi+\Gamma^{r}_{t t} (u^t)^2+\Gamma^{r}_{r r} (u^r)^2+\Gamma^{r}_{\phi \phi} (u^\phi)^2.
\end{align}

Using the four-velocity components from eqn. (\ref{4vel}) and $\disp{\Gamma^{r}_{t t} = \frac{M}{r^2} (1-2M/r)}$,\\ $\disp{\Gamma^{r}_{r r} = \frac{M}{r^2} (1-2M/r)^{-1}}$, $\disp{\Gamma^{r}_{\phi \phi} = -r \sin^2 \theta (1-2M/r)}$, 

%\begin{align}\label{ar}
%a^r &= \gamma^2 \left[\frac{M}{r^2}+\beta_r (1-2 M/r) (1+\beta^2_r \gamma^2) \frac{\partial \beta_r}{\partial r}\right.\\ \nonumber
%&\left.+\beta^2_r \gamma^2 (1-2 M/r) \left(\beta_\theta \frac{\partial \beta_\theta}{\partial r}+\beta_\phi \frac{\partial \beta_\phi}{\partial r}\right)-\frac{1}{r} (1-2 M/r) (\beta^2_\theta+\beta^2_\phi)\right. \\ \nonumber
%&\left.+\frac{(1-2 M/r)^{1/2}}{r} \left\{\beta_\theta \left((1+\beta^2_r \gamma^2) \frac{\partial\beta_r}{\partial \theta}+\beta_r \gamma^2 \left(\beta_\theta \frac{\partial\beta_\theta}{\partial \theta}+\beta_\phi \frac{\partial \beta_\phi}{\partial \theta}\right)\right)\right.\right.\\ \nonumber
%&\left.\left.+\frac{\beta_\phi}{\sin \theta} \left((1+\beta^2_r \gamma^2) \frac{\partial\beta_r}{\partial \phi}+\beta_r \gamma^2 \left(\beta_\theta \frac{\partial\beta_\theta}{\partial \phi}+\beta_\phi \frac{\partial \beta_\phi}{\partial \phi}\right)\right)\right\}\right]
%\end{align}

\begin{align}\label{ar}
a^r &= \gamma^2 \left[\frac{M}{r^2}+\beta_r (1-2 M/r) (1+\beta^2_r \gamma^2) \frac{\partial  \beta_r}{\partial r}+\beta^2_r \gamma^2 (1-2 M/r) \left(\beta_\theta \frac{\partial  \beta_\theta}{\partial r}+\beta_\phi \frac{\partial  \beta_\phi}{\partial r}\right)\right.\\ \nonumber
&\left.-\frac{M}{r} (1-2 M/r) (\beta^2_\theta+\beta^2_\phi) +\frac{\beta_\theta(1-2 M/r)^{1/2}}{r} \left\{(1+\beta^2_r \gamma^2) \frac{\partial\beta_r}{\partial \theta}+\beta_r \gamma^2 \left(\beta_\theta \frac{\partial\beta_\theta}{\partial \theta}+\beta_\phi \frac{\partial \beta_\phi}{\partial \theta}\right)\right\}\right.\\ \nonumber
&\left.+\frac{\beta_\phi (1-2 M/r)^{1/2}}{r \sin \theta} \left\{(1+\beta^2_r \gamma^2) \frac{\partial\beta_r}{\partial \phi}+\beta_r \gamma^2 \left(\beta_\theta \frac{\partial\beta_\theta}{\partial \phi}+\beta_\phi \frac{\partial \beta_\phi}{\partial \phi}\right)\right\}\right]
\end{align}

The quantity $m c^2 a^r$ is the radial force where $m$ is the mass of the orbiting blob. The $\theta$-component of the acceleration is given by
\begin{align}
a^\theta &= \frac{\partial u^\theta}{\partial r} u^r+\frac{\partial u^\theta}{\partial \theta} u^\theta+\frac{\partial u^\theta}{\partial \phi} u^\phi+\Gamma^{\theta}_{\mu \nu} u^{\mu} u^{\nu}\\ \nonumber
&= \frac{\partial u^\theta}{\partial r} u^r+\frac{\partial u^\theta}{\partial \theta} u^\theta+\frac{\partial u^\theta}{\partial \phi} u^\phi+\Gamma^{\theta}_{\phi \phi} (u^\phi)^2+2 \Gamma^{\theta}_{r \theta} u^r u^\theta.
\end{align}

Using the four-velocity components from eqn. (\ref{4vel}), $\Gamma^{\theta}_{\phi \phi} = -\cos \theta \sin \theta$ and $\Gamma^{\theta}_{r \theta} = 1/r$,

%\begin{align}\label{atheta}
%a^\theta &= \frac{\gamma^2}{r} \left[\beta_r (1-2M/r)^{1/2} \left\{(1+\beta^2_\theta \gamma^2) \frac{\partial \beta_\theta}{\partial r} \right.\right.\\ \nonumber
%&\left.\left.+\beta_\theta \gamma^2 \left(\beta_r \frac{\partial \beta_r}{\partial r}+\beta_\phi \frac{\partial \beta_\phi}{\partial r}\right)-\frac{\beta_\theta}{r}\right\}-\frac{\cot \theta}{r} \beta^2_\phi+\frac{2}{r} (1-2 M/r)^{1/2} \beta_r \beta_\theta \right. \\ \nonumber
%&+\left.\frac{\beta_\theta}{r} \left((1+\beta^2_\theta \gamma^2) \frac{\partial\beta_\theta}{\partial \theta}+\beta_\theta \gamma^2 \left(\beta_r \frac{\partial\beta_\theta}{\partial \theta}+\beta_\phi \frac{\partial \beta_\phi}{\partial \theta}\right)\right)\right.\\ \nonumber
%&\left.+\frac{\beta_\phi}{r \sin \theta} \left((1+\beta^2_\theta \gamma^2) \frac{\partial\beta_\theta}{\partial \phi}+\beta_\theta \gamma^2 \left(\beta_r \frac{\partial\beta_r}{\partial \phi}+\beta_\phi \frac{\partial \beta_\phi}{\partial \phi}\right)\right) \right],
%\end{align}

\begin{align}\label{atheta}
a^\theta &= \frac{\gamma^2}{r} \left[\beta_r (1-2 M/r)^{1/2} \left\{(1+\beta^2_\theta \gamma^2) \frac{\partial \beta_\theta}{\partial r}+\beta_\theta \gamma^2 \left(\beta_r \frac{\partial \beta_r}{\partial r}+\beta_\phi \frac{\partial \beta_\phi}{\partial r}\right)+\frac{\beta_\theta}{r}\right\}\right.\\ \nonumber
&\left.+\frac{\beta_\theta}{r} \left\{(1+\beta^2_\theta \gamma^2) \frac{\partial\beta_\theta}{\partial \theta}+\beta_\theta \gamma^2 \left(\beta_r \frac{\partial\beta_r}{\partial \theta}+\beta_\phi \frac{\partial \beta_\phi}{\partial \theta}\right)\right\}\right.\\ \nonumber
&\left.+\frac{\beta_\phi}{r \sin \theta} \left\{(1+\beta^2_\theta \gamma^2) \frac{\partial\beta_\theta}{\partial \phi}+\beta_\theta \gamma^2 \left(\beta_r \frac{\partial\beta_r}{\partial \phi}+\beta_\phi \frac{\partial \beta_\phi}{\partial \phi}\right)\right\}-\frac{\cot \theta}{r} \beta^2_\phi\right]
\end{align}

The quantity $m c^2 r a^\theta$ is the force in the $\theta$ direction. The azimuthal component is given by
\begin{align}
a^\phi &= \frac{\partial u^\phi}{\partial r} u^r+\frac{\partial u^\phi}{\partial \theta} u^\theta+\frac{\partial u^\phi}{\partial \phi} u^\phi+\Gamma^{\phi}_{\mu \nu} u^{\mu} u^{\nu}\\ \nonumber
&= \frac{\partial u^\phi}{\partial r} u^r+\frac{\partial u^\phi}{\partial \theta} u^\theta+\frac{\partial u^\phi}{\partial \phi} u^\phi+2 \Gamma^{\phi}_{r \phi} u^r u^\phi+2 \Gamma^{\phi}_{\theta \phi} u^\theta u^\phi.
\end{align}

Using the four-velocity components from eqn. (\ref{4vel}), $\Gamma^{\phi}_{r \phi} = 1/r$ and $\Gamma^{\phi}_{\theta \phi} = \cot \theta$, 

%\begin{align}\label{aphi}
%a^\phi &= \frac{1}{r \sin \theta} \beta_r \gamma^2 (1-2M/r)^{1/2} \left[(1+\beta^2_\phi \gamma^2) \frac{\partial \beta_\phi}{\partial r}+\frac{\beta_\phi}{r}-\beta_\phi \cot \theta \frac{\partial \beta_\theta}{\partial r}\right. \\ \nonumber
%&\left.+\beta_\phi \gamma^2 \left(\beta_r \frac{\partial \beta_r}{\partial r}+\beta_\theta \frac{\partial \beta_\theta}{\partial r}\right)+\frac{2}{r \beta_r} (1-2 M/r)^{-1/2} \beta_\theta \beta_\phi \cot \theta \right. \\ \nonumber
%&\left.+\frac{(1-2 M/r)^{-1/2}}{r \beta_r} \left\{\beta_\theta \left((1+\beta^2_\phi \gamma^2) \frac{\partial\beta_\phi}{\partial \theta}+\beta_\phi \gamma^2 \left(\beta_r \frac{\partial\beta_r}{\partial \theta}+\beta_\theta \frac{\partial \beta_\theta}{\partial \theta}\right)\right)\right.\right.\\ \nonumber
%&\left.\left.+\frac{\beta_\phi}{\sin \theta} \left((1+\beta^2_\phi \gamma^2) \frac{\partial\beta_\phi}{\partial \phi}+\beta_\phi \gamma^2 \left(\beta_r \frac{\partial\beta_r}{\partial \phi}+\beta_\theta \frac{\partial \beta_\theta}{\partial \phi}\right)\right)\right\}\right]
%\end{align}

\begin{align*}\label{aphi}
a^\phi &= \frac{\gamma^2}{r \sin \theta} \left[\beta_r (1-2M/r)^{1/2}\left\{(1+\beta^2_\phi \gamma^2) \frac{\partial \beta_\phi}{\partial r}+\beta_\phi\left(\frac{M}{r}-\cot \theta \frac{\partial \beta_\theta}{\partial r}+\gamma^2 \left(\beta_r \frac{\partial \beta_r}{\partial r}+\beta_\theta \frac{\partial \beta_\theta}{\partial r}\right)\right)\right\}\right. \\ \nonumber 
&\left.+\frac{\beta_\theta}{r} \left\{(1+\beta^2_\phi \gamma^2) \frac{\partial\beta_\phi}{\partial \theta}+2 \beta_\phi \cot \theta+\beta_\phi \gamma^2 \left(\beta_r \frac{\partial\beta_r}{\partial \theta}+\beta_\theta \frac{\partial \beta_\theta}{\partial \theta}\right)\right\}\right.\\ \nonumber
&\left.+\frac{\beta_\phi}{r \sin \theta} \left\{(1+\beta^2_\phi \gamma^2) \frac{\partial\beta_\phi}{\partial \phi}+\beta_\phi \gamma^2 \left(\beta_r \frac{\partial\beta_r}{\partial \phi}+\beta_\theta \frac{\partial \beta_\theta}{\partial \phi}\right)\right\}\right].\numberthis
\end{align*}

The quantity $m c^2 r a^\phi$  represents the force in azimuthal direction. Radiation force drives the accelerated outward motion of the blob while the initial azimuthal and angular motion however is continuously retarded by the effect of radiation drag of an assumed spherically symmetric radiation field.

Following \cite{1990ApJ...361..470A}, the contra-variant components of the radiation energy flux are given by,
\begin{equation}
F^\alpha = h^{\alpha}_{\mu} T^{\mu \nu} u_{\nu},
\end{equation}
where $h^{\alpha}_{\mu} = \delta^{\alpha}_{\mu}+u^{\alpha} u_{\mu}$ is a projection tensor and $T^{\mu \nu}$ is the radiation field energy density. The tetrad components of the energy density tensor can be evaluated using 
\begin{equation}
T^{(i)(k)} = \int I(r) n^{(i)} n^{(k)} d \Omega
\end{equation}
where $I(r)$ is the radiation field intensity, $n^{(i)}$ are the unit vectors describing the photon trajectory and $d \Omega$ is an element of the solid angle subtended on the sky of the local observer (see Fig. 3 of \citealt{1990ApJ...361..470A}). The energy density can thus be evaluated from this using 
\begin{align}\label{enden}
T^{tt} &= T^{(t)(t)} (1-2M/r)^{-1} = 2 \pi I(r) (1-2M/r)^{-1} (1-\cos \eta)\\ \nonumber
T^{tr} &= T^{rt}= T^{(t)(r)}= T^{(r)(t)} = \pi I(r) \sin^2 \eta\\ \nonumber
T^{rr} &= T^{(r)(r)} (1-2 M/r) = \frac{2 \pi I(r)}{3} (1-2 M/r) (1-\cos^3\eta)\\ \nonumber
T^{\theta \theta} &= \frac{T^{(\theta)(\theta)}}{r^2} = \frac{\pi I(r)}{3 r^2} (\cos^3\eta-3 \cos \eta+2)\\ \nonumber
T^{\phi \phi} &= \frac{T^{(\phi)(\phi)}}{r^2 \sin^2 \theta} = \frac{\pi I(r)}{3 r^2 \sin^2 \theta} (\cos^3\eta-3 \cos \eta+2),
\end{align}
where $R = 6 M$ is the radius of the emitting surface, $\eta$ is the viewing angle from the zenith for the stationary observer and is related to $\delta$, the zenithal angle at the point of emission, $R$ by
\begin{equation}
\sin \eta = \frac{M}{r} \frac{(1-2M/r)^{1/2}}{(1-2M/R)^{1/2}} \sin \delta,
\end{equation}
where $\sin \delta = 1$ for $R/M > 3/2$ and $\disp{\sin \delta = \frac{3 \sqrt{3}}{2} \frac{M}{R} (1-2M/R)^{1/2}}$ for $R/M \leq 3/2$ (\citealt{1990ApJ...361..470A}).  

If $\sigma$ is the cross section of the orbiting blob over which the radiation force acts, the quantity $\sigma F^\alpha$ represents the radiation force. The radial component of the energy flux is given by
\begin{equation}
F^r = [(1+u^r u_r) (T^{r t} u_t+T^{r r} u_r)+u^r u_t (T^{t t} u_t+T^{t r} u_r)+u^r u_{\theta} (T^{\theta \theta} u_{\theta})+u^r u_{\phi} (T^{\phi \phi} u_{\phi})].
\end{equation}

Using the components of $u^{\alpha}$ from eqn. (\ref{4vel}), $u_{\alpha}$ from eqn. (\ref{4velc}) and $T^{\mu \nu}$ from eqn. (\ref{enden}), the radiation force in the radial direction is given by

\begin{align}\label{Fr}
\sigma F^r &= \pi I(r) \sigma \gamma^3 (1-2M/r)^{1/2} \left[\frac{1+\beta^2_r \gamma^2}{\gamma^2} \left(-\sin^2 \eta +\frac{2}{3} \beta_r (1-\cos^3 \eta)\right)\right.\\ \nonumber
&\left.-\beta_r (\beta_r \sin^2 \eta-2 (1-\cos \eta))+\frac{\beta_r}{3} (\beta^2_\theta+\beta^2_\phi) (\cos^3 \eta-3 \cos \eta+2)\right].
\end{align}

The $\theta$-component of the energy flux is given by
\begin{equation}
F^\theta = [(1+u^\theta u_\theta) T^{\theta \theta} u_\theta+u^\theta u_t (T^{t t} u_t+T^{t r} u_r)+u^\theta u_r (T^{r t} u_t+T^{r r} u_r)+u^\theta u_\phi T^{\phi \phi} u_\phi].
\end{equation}

Using the components of $u^{\alpha}$ from eqn. (\ref{4vel}), $u_{\alpha}$ from eqn. (\ref{4velc}) and the components of $T^{\mu \nu}$ from eqn. (\ref{enden}), the $\theta$-component of the radiation force is given by
\begin{align}\label{Ftheta}
\sigma F^\theta &= \frac{\pi I(r) \sigma \gamma^3 \beta_\theta}{r} \left[\frac{1+\beta^2_\theta \gamma^2}{3 \gamma^2} (\cos^3 \eta-3 \cos \eta+2) \right. \\ \nonumber
&\left.+ 2 (1-\cos \eta)-\frac{2}{3} \beta^2_r (1-\cos^3 \eta)+\beta^2_\phi (\cos^3 \eta-3 \cos \eta+2)\right].
\end{align}

The azimuthal component of the energy flux is given by
\begin{equation}
F^\phi = [(1+u^\phi u_\phi) T^{\phi \phi} u_\phi+u^\phi u_t (T^{t t} u_t+T^{t r} u_r)+u^\phi u_r (T^{r t} u_t+T^{r r} u_r)+u^\phi u_\theta T^{\theta \theta} u_\theta].
\end{equation}

Using the components of $u^{\alpha}$ from eqn. (\ref{4vel}), $u_{\alpha}$ from eqn. (\ref{4velc}) and the components of $T^{\mu \nu}$ from eqn. (\ref{enden}), the azimuthal component of the radiation force is given by
\begin{align}\label{Fphi}
\sigma F^\phi &= \frac{\pi I(r) \sigma \gamma^3 \beta_\phi}{r \sin \theta} \left[\frac{1+\beta^2_\phi \gamma^2}{3 \gamma^2} (\cos^3 \eta-3 \cos \eta+2) \right. \\ \nonumber
&\left.+ 2 (1-\cos \eta)-\frac{2}{3} \beta^2_r (1-\cos^3 \eta)+\beta^2_\theta (\cos^3 \eta-3 \cos \eta+2)\right].
\end{align}

\subsection{Dynamics based on radiation pressure and drag}
\label{radprdynamics}

The Eddington parameter $\Gamma_{\epsilon}$ is the ratio of the disk luminosity $L$ to the Eddington luminosity $L_\mathrm{Edd}$ and is given by, 
\begin{equation}
\Gamma_{\epsilon} = \frac{L}{L_\mathrm{Edd}} = \frac{c \pi I(R) \sigma  R^2 (1-2M/R)^{1/2}}{G M_{\bullet} m c}
\label{Eddpar}
\end{equation}

where $L_\mathrm{Edd}$ is corrected for the general relativistic redshift factor $(1-2M/R)^{1/2}$ and $\sigma$ is the cross section of the orbiting blob over which the radiation force acts and
\begin{equation}
\frac{I(r)}{I(R)} = \left(\frac{1-2M/R}{1-2M/r}\right)^2,
\label{IrIR}
\end{equation}
is obtained from the energy conservation along the null trajectory, in the stationary frame (\citealt{1990ApJ...361..470A}).

The first equation of motion is from the radial components of the acceleration and the force. The radial component of the force imparts the radial acceleration onto the orbiting blob, i.e. $\sigma F^r = m c^2 a^r$. Using eqns. (\ref{ar}) and (\ref{Fr}) and the representation $x=r/M$ and $X=R/M$, we can eliminate the dependence of the equation on $M$. Then, expressing the equation in terms of the Eddington parameter $\Gamma_{\epsilon}$ using eqns. (\ref{Eddpar}) and (\ref{IrIR}),
%\begin{align}\label{em1}
%&\frac{1}{x^2}+\beta_r (1-2/x) (1+\beta^2_r \gamma^2) \frac{\partial  \beta_r}{\partial x}+\beta^2_r \gamma^2 (1-2/x) \left(\beta_\theta \frac{\partial  \beta_\theta}{\partial x}+\beta_\phi \frac{\partial  \beta_\phi}{\partial x}\right)\\ \nonumber
%&-\frac{1}{x} (1-2/x) (\beta^2_\theta+\beta^2_\phi) +\frac{(1-2/x)^{1/2}}{x} \left\{\beta_\theta \left((1+\beta^2_r \gamma^2) \frac{\partial\beta_r}{\partial \theta}+\beta_r \gamma^2 \left(\beta_\theta \frac{\partial\beta_\theta}{\partial \theta}+\beta_\phi \frac{\partial \beta_\phi}{\partial \theta}\right)\right)\right.\\ \nonumber
%&\left.+\frac{\beta_\phi}{\sin \theta} \left((1+\beta^2_r \gamma^2) \frac{\partial\beta_r}{\partial \phi}+\beta_r \gamma^2 \left(\beta_\theta \frac{\partial\beta_\theta}{\partial \phi}+\beta_\phi \frac{\partial \beta_\phi}{\partial \phi}\right)\right)\right\}\\ \nonumber
%&=\frac{\Gamma_{\epsilon} \gamma}{X^2} \left(\frac{1-2/X}{1-2/x}\right)^{3/2} \left[\frac{1+\beta^2_r \gamma^2}{\gamma^2} \left(-\sin^2 \eta +\frac{2}{3} \beta_r (1-\cos^3 \eta)\right)\right.\\ \nonumber
%&\left.-\beta_r (\beta_r \sin^2 \eta-2 (1-\cos \eta))+\frac{\beta_r}{3} (\beta^2_\theta+\beta^2_\phi) (\cos^3 \eta-3 \cos \eta+2)\right].  
%\end{align}

\begin{align}\label{em1}
&\frac{1}{x^2}+\beta_r (1-2/x) (1+\beta^2_r \gamma^2) \frac{\partial  \beta_r}{\partial x}+\beta^2_r \gamma^2 (1-2/x) \left(\beta_\theta \frac{\partial  \beta_\theta}{\partial x}+\beta_\phi \frac{\partial  \beta_\phi}{\partial x}\right)\\ \nonumber
&-\frac{1}{x} (1-2/x) (\beta^2_\theta+\beta^2_\phi) +\frac{\beta_\theta(1-2/x)^{1/2}}{x} \left\{(1+\beta^2_r \gamma^2) \frac{\partial\beta_r}{\partial \theta}+\beta_r \gamma^2 \left(\beta_\theta \frac{\partial\beta_\theta}{\partial \theta}+\beta_\phi \frac{\partial \beta_\phi}{\partial \theta}\right)\right\}\\ \nonumber
&+\frac{\beta_\phi (1-2/x)^{1/2}}{x \sin \theta} \left\{(1+\beta^2_r \gamma^2) \frac{\partial\beta_r}{\partial \phi}+\beta_r \gamma^2 \left(\beta_\theta \frac{\partial\beta_\theta}{\partial \phi}+\beta_\phi \frac{\partial \beta_\phi}{\partial \phi}\right)\right\}\\ \nonumber
&=\frac{\Gamma_{\epsilon} \gamma}{X^2} \left(\frac{1-2/X}{1-2/x}\right)^{3/2} \left[\frac{1+\beta^2_r \gamma^2}{\gamma^2} \left(-\sin^2 \eta +\frac{2}{3} \beta_r (1-\cos^3 \eta)\right)\right.\\ \nonumber
&\left.-\beta_r \left(\beta_r \sin^2 \eta-2 (1-\cos \eta)+\frac{\beta^2_\theta+\beta^2_\phi}{3} (\cos^3 \eta-3 \cos \eta+2)\right)\right].  
\end{align}

%\begin{align}
%&\frac{1}{x^2}+\beta_r (1-2/x) (1+\beta^2_r \gamma^2) \frac{\partial  \beta_r}{\partial x}+\beta_\theta (1-2/x)^{1/2} \left\{\beta^2_r \gamma^2 (1-2/x)^{1/2} \left(\beta_\theta \frac{\partial  \beta_\theta}{\partial x}+\beta_\phi \frac{\partial  \beta_\phi}{\partial x}\right)\right. \\ \nonumber
%&\left.+\frac{1}{x} \left((1+\beta^2_r \gamma^2) \frac{\partial\beta_r}{\partial \theta}+\beta_r \gamma^2 \left(\beta_\theta \frac{\partial\beta_\theta}{\partial \theta}+\beta_\phi \frac{\partial \beta_\phi}{\partial \theta}\right)\right)\right\}
%\end{align}

The second equation of motion is from the $\theta$-components of the acceleration and the force. The $\theta$-component of the force imparts a drag on the acceleration of the orbiting blob and $\sigma F^\theta = m c^2 a^\theta$.  Using $\gamma_r = (1-\beta^2_r)^{1/2}$, eqns. (\ref{atheta}) and (\ref{Ftheta}) and the representation $x=r/M$ and $X=R/M$ and expressing the equation in terms of the Eddington parameter $\Gamma_{\epsilon}$ using eqns. (\ref{Eddpar}) and (\ref{IrIR}),
\begin{align}\label{em2}
&\beta_r (1-2/x)^{1/2} \left\{(1+\beta^2_\theta \gamma^2) \frac{\partial \beta_\theta}{\partial x}+\beta_\theta \gamma^2 \left(\beta_r \frac{\partial \beta_r}{\partial x}+\beta_\phi \frac{\partial \beta_\phi}{\partial x}\right)+\frac{\beta_\theta}{x}\right\}\\ \nonumber
&+\frac{\beta_\theta}{x} \left\{(1+\beta^2_\theta \gamma^2) \frac{\partial\beta_\theta}{\partial \theta}+\beta_\theta \gamma^2 \left(\beta_r \frac{\partial\beta_r}{\partial \theta}+\beta_\phi \frac{\partial \beta_\phi}{\partial \theta}\right)\right\}\\ \nonumber
&+\frac{\beta_\phi}{x \sin \theta} \left\{(1+\beta^2_\theta \gamma^2) \frac{\partial\beta_\theta}{\partial \phi}+\beta_\theta \gamma^2 \left(\beta_r \frac{\partial\beta_r}{\partial \phi}+\beta_\phi \frac{\partial \beta_\phi}{\partial \phi}\right)\right\}-\frac{\cot \theta}{x} \beta^2_\phi\\ \nonumber
&= \frac{\Gamma_{\epsilon} \gamma \beta_\theta}{X^2 (1-2/X)^{1/2}} \left(\frac{1-2/X}{1-2/x}\right)^{2} \\ \nonumber
&\left[2 (1-\cos \eta)-\frac{2}{3} \beta^2_r (1-\cos^3 \eta)+ \frac{1}{3}\left(\frac{1}{\gamma^2_r}+2 \beta^2_\phi\right) (\cos^3 \eta-3 \cos \eta+2)\right].
\end{align}

The third equation of motion is from the azimuthal components of the acceleration and the force. The azimuthal component of the force imparts a drag on the azimuthal acceleration of the orbiting blob and $\sigma F^\phi = m c^2 a^\phi$. Using eqns. (\ref{aphi}) and (\ref{Fphi}) and the representation $x=r/M$ and $X=R/M$ and expressing the equation in terms of the Eddington parameter $\Gamma_{\epsilon}$ using eqns. (\ref{Eddpar}) and (\ref{IrIR}),
\begin{align}\label{em3}
&\beta_r (1-2/x)^{1/2} \left\{ (1+\beta^2_\phi \gamma^2) \frac{\partial \beta_\phi}{\partial x}+\beta_\phi\left(\frac{1}{x}-\cot \theta \frac{\partial \beta_\theta}{\partial x}+\gamma^2 \left(\beta_r \frac{\partial \beta_r}{\partial x}+\beta_\theta \frac{\partial \beta_\theta}{\partial x}\right)\right)\right\} \\ \nonumber 
&+\frac{\beta_\theta}{x} \left\{(1+\beta^2_\phi \gamma^2) \frac{\partial\beta_\phi}{\partial \theta}+2 \beta_\phi \cot \theta+\beta_\phi \gamma^2 \left(\beta_r \frac{\partial\beta_r}{\partial \theta}+\beta_\theta \frac{\partial \beta_\theta}{\partial \theta}\right)\right\}\\ \nonumber
&+\frac{\beta_\phi}{x \sin \theta} \left\{(1+\beta^2_\phi \gamma^2) \frac{\partial\beta_\phi}{\partial \phi}+\beta_\phi \gamma^2 \left(\beta_r \frac{\partial\beta_r}{\partial \phi}+\beta_\theta \frac{\partial \beta_\theta}{\partial \phi}\right)\right\}\\ \nonumber
&= \frac{\Gamma_{\epsilon} \gamma \beta_\phi}{X^2 (1-2/X)^{1/2}} \left(\frac{1-2/X}{1-2/x}\right)^{2}\\ \nonumber
&\left[2 (1-\cos \eta)-\frac{2}{3} \beta^2_r (1-\cos^3 \eta)+ \frac{1}{3}\left(\frac{1}{\gamma^2_r}+2 \beta^2_\theta\right) (\cos^3 \eta-3 \cos \eta+2)\right].
\end{align}

%The solution of the set of three coupled partial differential equations will 

\subsubsection{Purely radial motion}

\begin{figure}[h!]
\centerline{\includegraphics[scale=0.15]{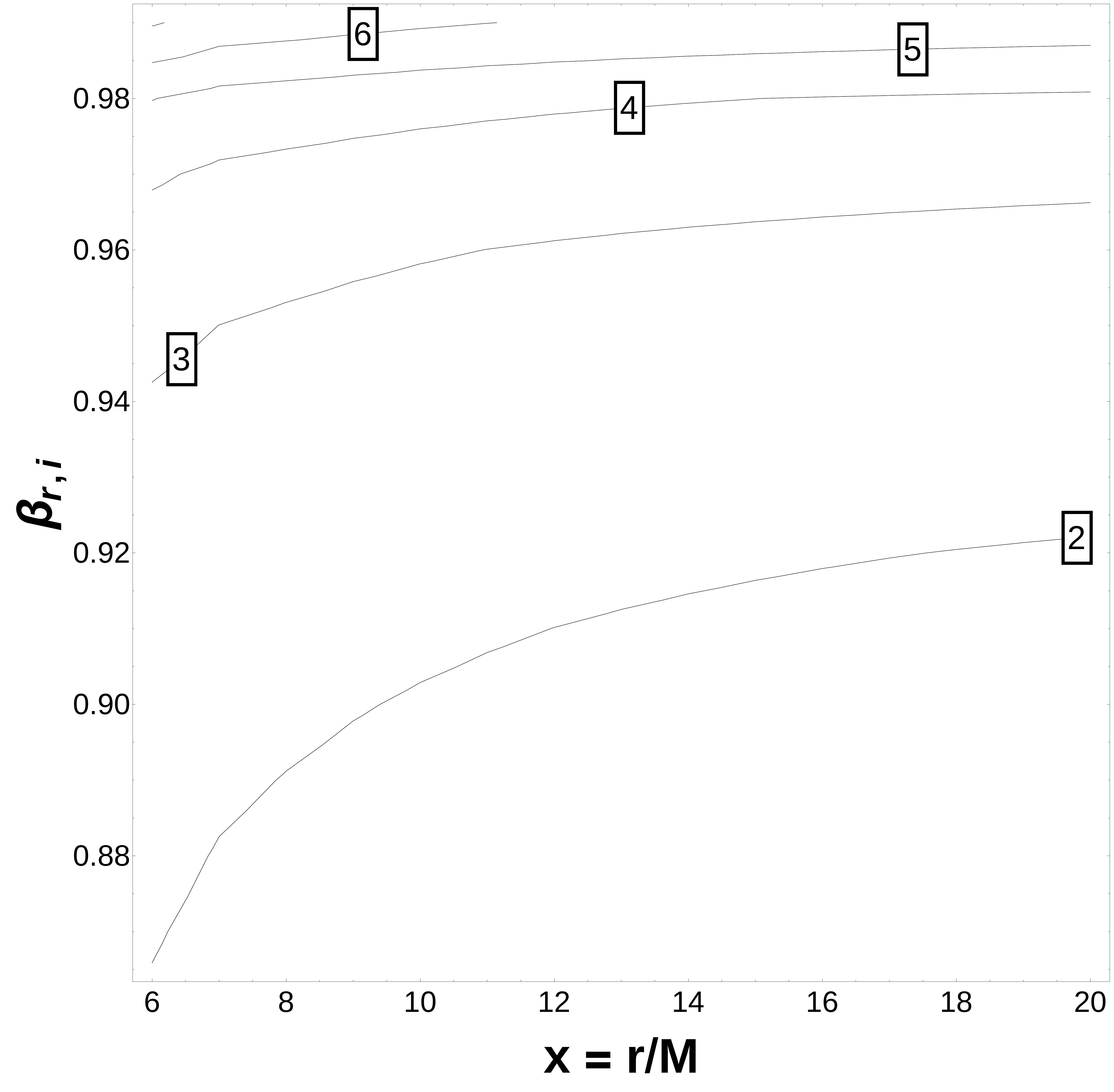}}
\caption{Contours of $\gamma_r$ as a function of the initial launch velocity $\beta_{r,i}$ and $x$. The $\gamma_r$ value tapers off to constant values at large $x$ for $\beta_{r,i} = 0.86 - 0.99$ consistent with the simulations performed in \cite{1990ApJ...361..470A}.
 The simulated $\gamma_r$ are in the range $2 - 7$.}
\label{gammarx}
\end{figure}

In the limit $\beta_\phi = 0$ and $\beta_\theta = 0$, i.e. purely radial outflowing motion, only the first equation of motion is relevant. The equation reduces to 
\begin{equation}
(1-2/x) \gamma^2_r \beta_r \frac{d\beta_r}{dx}+\frac{1}{x^2}= \frac{\Gamma_{\epsilon}}{x^2} \left(\frac{1-2/X}{1-2/x}\right)^{3/2} \left[-\sin^2 \eta (1+\beta^2_r)+\frac{2}{3} \beta_r (4-\cos^3 \eta-3 \cos \eta)\right],
\end{equation}
which is the same as that derived in (\citealt{1990ApJ...361..470A}). Solving for $\beta_r$ and hence $\gamma_r = (1-\beta^2_r)^{-1/2}$ as a function of the initial launch velocity, $\beta_{r,i}$ and $x$ indicates a constant saturation value of $2-7$ for $\Gamma_{\epsilon} = 0.3$, $\beta_{r,i} = 0.86 - 0.99$ at distances $r$ beyond $\sim$ a few tens of $M$ from the black hole and is plotted in Fig. \ref{gammarx}. In the simulations, changes to $\Gamma_{\epsilon}$ did not change these results. 

For the range of $\beta_{r,i}= 0.86 - 0.99$ and $x = 6 - 20$, the $\gamma_r$ contours saturate at constant values in the range $2 - 7$ for large $x$, consistent with the simulations of \cite{1990ApJ...361..470A}.

\subsubsection{Poloidal motion}

%The study of the general case involves solving the radial, $\theta$ and azimuthal equations, eqns. (\ref{em1}), (\ref{em2}) and (\ref{em3}) simultaneously. 
A poloidal outflow in the radiation region ($\beta_\phi = 0, \beta_\theta\neq 0, \beta_r \neq 0$) is trajectorially equivalent to the case of motion on the equatorial plane ($\theta = \pi/2, \beta_\theta = 0, \beta_\phi \neq 0, \beta_r \neq 0$) due to spherical symmetry of the Schwarzschild geometry.  We appeal to the zero angular momentum nature of the outflow source in the corona region (see Fig. \ref{RadiationGeometry}) to motivate this poloidal flow. Thus, we are left with two coupled partial differential equations to solve, modified versions of eqns. (\ref{em1}) and (\ref{em2}) with ($\beta_\phi = 0, \beta_\theta\neq 0, \beta_r \neq 0$). The bulk Lorentz factor $\gamma = (1-\beta^2_r-\beta^2_\theta)^{-1/2}$ in this case and $\gamma_r = (1-\beta^2_r)^{-1/2}$; $\gamma_\theta = (1-\beta^2_\theta)^{-1/2}$. The resulting equations  are,
\begin{align}
&\frac{1}{x^2}+\beta_r \gamma^2 (1-2/x) \left(\frac{1}{\gamma^2_\theta} \frac{\partial \beta_r}{\partial x}+\beta_r \beta_\theta \frac{\partial  \beta_\theta}{\partial x}\right)-\frac{\beta^2_\theta}{x} (1-2/x)\\ \nonumber
& +\frac{\beta_\theta \gamma^2}{x} (1-2/x)^{1/2} \left(\frac{1}{\gamma^2_\theta}\frac{\partial\beta_r}{\partial \theta}+\beta_r \beta_\theta \frac{\partial\beta_\theta}{\partial \theta}\right)\\ \nonumber\\ \nonumber
&=\frac{\Gamma_{\epsilon} \gamma}{X^2} \left(\frac{1-2/X}{1-2/x}\right)^{3/2} \left[\frac{1}{\gamma^2_\theta} \left(-\sin^2 \eta +\frac{2}{3} \beta_r (1-\cos^3 \eta)\right)\right.\\ \nonumber
&\left.-\beta_r \left(\beta_r \sin^2 \eta-2 (1-\cos \eta)+\frac{\beta^2_\theta}{3} (\cos^3 \eta-3 \cos \eta+2)\right)\right].  
\end{align}
for the radial motion and the $\theta$-equation is,
\begin{align}
&\beta_r (1-2/x)^{1/2} \left(\frac{\gamma^2}{\gamma^2_r} \frac{\partial \beta_\theta}{\partial x}+\beta_\theta \beta_r \gamma^2 \frac{\partial \beta_r}{\partial x}+\frac{\beta_\theta}{x}\right)+\frac{\beta_\theta \gamma^2}{x} \left(\frac{1}{\gamma^2_r} \frac{\partial\beta_\theta}{\partial \theta}+\beta_\theta \beta_r \frac{\partial\beta_r}{\partial \theta}\right)\\ \nonumber
&= \frac{\Gamma_{\epsilon} \gamma \beta_\theta}{X^2 (1-2/X)^{1/2}} \left(\frac{1-2/X}{1-2/x}\right)^{2} \left[2 (1-\cos \eta)-\frac{2}{3} \beta^2_r (1-\cos^3 \eta)+\frac{1}{3 \gamma^2_r} (\cos^3 \eta-3 \cos \eta+2)\right].
\end{align}

In the simulations carried out, we set $\Gamma_{\epsilon} = 0.3$. The initial value $\beta_{r,i}$ was varied between 0.01 - 0.99, the range being chosen based on the $\beta_r$ range simulated in (\citealt{2014ApJ...791...74M}) for a relativistic thin disk (\citealt{1973A&A....24..337S,1973blho.conf..343N}) in the context of the quality factor $Q$ observable in the inner region of the accretion disk very close to the innermost stable circular orbit. There, we obtained the range $1.36 \times 10^5 - 0.99$ for a range of the disk viscosity parameter $\alpha = 0.01 - 0.4$ for $r \geq 6 M - 20 M$. The range of initial velocity $\beta_{\theta,i}$ is chosen based on the argument that if the motion is along Keplerian orbits, $\disp{\beta_\theta = (1-2 M/r)^{-1/2} \frac{r}{c} \frac{d\theta}{dt} = (1-2 M/r)^{-1/2} \frac{r}{c} \Omega(r)}$. As $\Omega(r) = c (M/r^3)^{1/2}$ for Keplerian orbits, $\beta_\theta =(1-2 M/r)^{-1/2} (M/r)^{1/2}$. We thus evaluate $\beta_{\theta,i}$ to be in the range $0.25 - 0.61$ for $r = 6 M - 20 M$. 

\begin{figure}[!h]
\centerline{\includegraphics[scale=0.15]{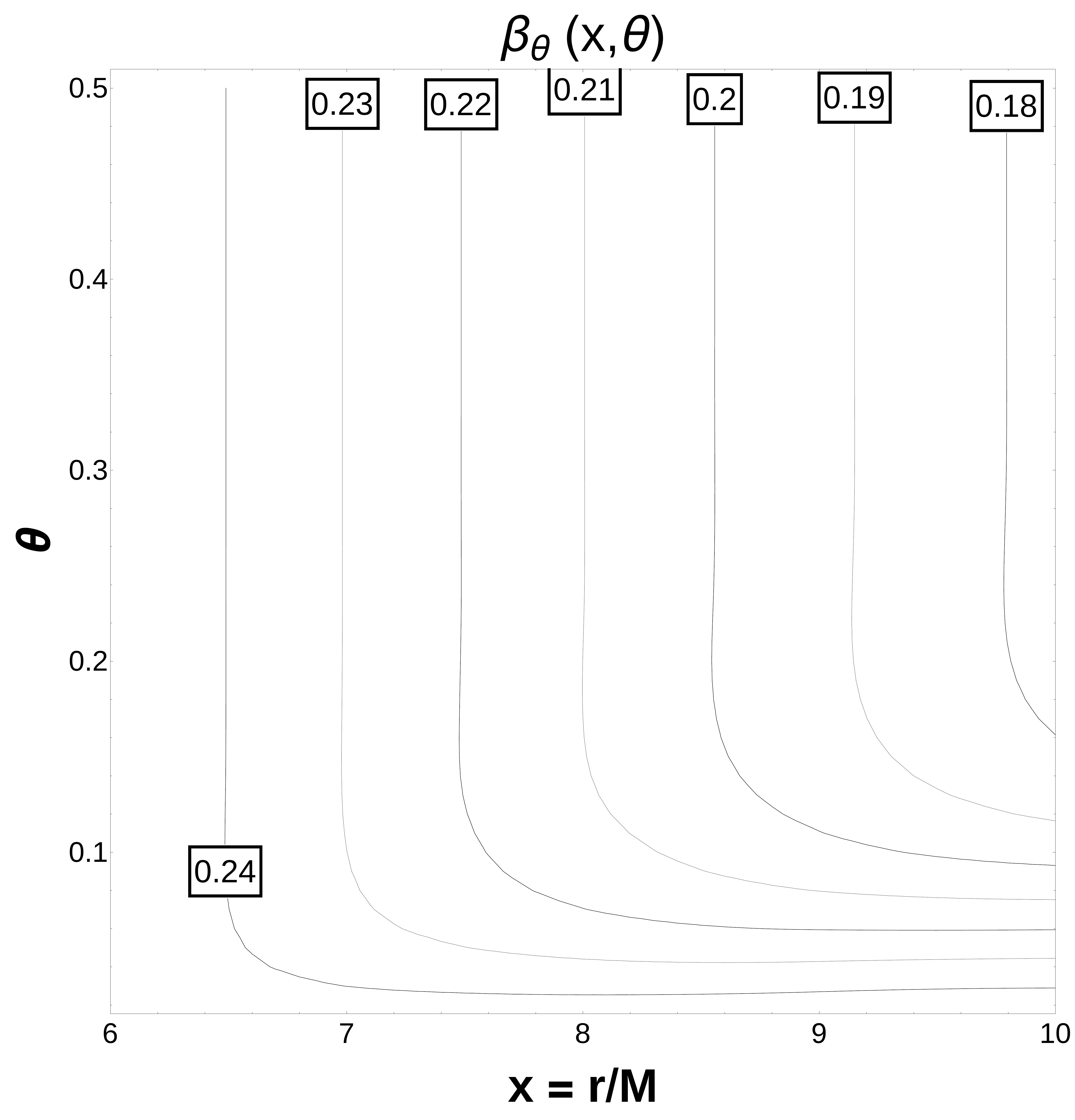}\includegraphics[scale=0.15]{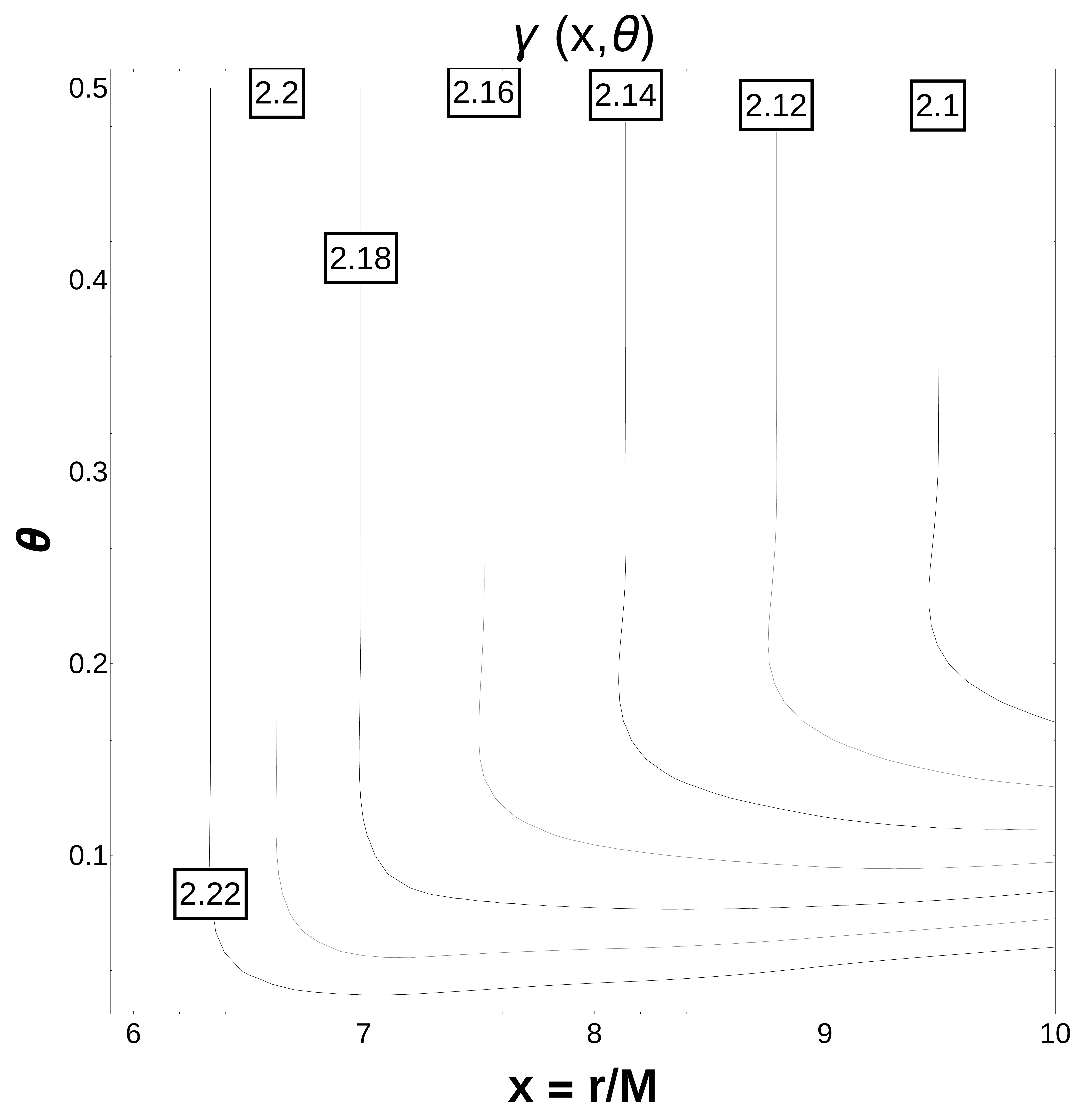}}
\caption{Contour plots of $\beta_\theta (x,\theta)$ and $\gamma (x,\theta)= (1-\beta^2)^{-1/2}$ where $\beta = (\beta^2_r+\beta^2_\theta)^{1/2}$ for $\beta_{r,i} = 0.86 - 0.99$ and $\beta_{\theta,i} = 0.25 - 0.61$. Left plot: simulated $\beta_\theta$ are in the range $0.18 - 0.81$. The decrease in $\beta_\theta$ is large for smaller initial values of $x$. Right plot: the decrease in $\gamma$ also follows the same expected trend to that of $\beta_\theta$. The simulated $\gamma$ are in the range $1.1 - 26.3$ higher than in the purely radial case though with saturation of $\gamma$ to low values occurring even at smaller $x$. These effects can be attributed to the drag force acting to rapidly cause the loss of any initial angular momentum as well as stabilize the flow at small $x$.}
\label{betaGammaxtheta}
\end{figure}

The initial values $\beta_{r,i}$ and $\beta_{\theta,i}$ were taken in the range $0.86 - 0.99$ (similar to the case of purely radial flow) and $0.25 - 0.61$ respectively. The initial values of the variables $x$ and $\theta$ were taken in the range $6 - 20$ and $0.01 - \pi/2$ respectively. The contours of $\beta_\theta$ and $\gamma = (1-\beta^2)^{-1/2}$ where $\beta = (\beta^2_r+\beta^2_\theta)^{1/2}$ as a function of $x$ and $\theta$ are plotted in Fig. \ref{betaGammaxtheta}. Simulated final values of $\beta_\theta$ are in the range $0.18 - 0.81$ and $\gamma$ are in the range $1.1 - 26.3$, higher than in the purely radial case though, $\gamma$ saturates to low values for smaller initial values of $x$. It is inferred from the simulations that $\beta_\theta$ and $\gamma$ decrease rapidly; this decrease is large for smaller initial values of $x$, while their decrease is more gradual at larger initial values of $x$. These effects can be attributed to the drag force due to the azimuthal component of the radiation pressure which plays an important role in stabilizing the saturation values of $\gamma$ at an early stage even if the source of the outflow had an initial angular momentum. The simulations thus indicate that the flow becomes rapidly radial.

We plan to present details of classification of flows and hence, a more comprehensive exploration of the parameter space and the resulting trajectories in a paper in preparation. Here, we have derived the typical values of the poloidal $\beta_p$ and $\gamma_p$ that are the inputs to typical initial values for Zone 3 ($R_A < r < R_L$) and beyond ($r > R_L$) where the angular momentum is set by its value at the Alfv\'{e}n point.  

\section{Trajectory and emission geometry in and beyond Zone 3}
\label{tframework}

From the metric in eqn. (\ref{metric}), as there is no explicit dependence in the equations of motion on the $\phi$ and $t$ coordinates, there are two Killing vectors associated with this geometry given by ${\bf \zeta} = (1,0,0,0)$ and ${\bf \eta} = (0,0,0,1)$. If $u$ is the four-velocity of the emitting blob, the Killing vectors can be used to evaluate the constants of motion for the Schwarzschild metric including the the total energy
\begin{equation}
\varepsilon = -{\bf \zeta \cdot u} = \frac{u^t}{(1-2 M/R)},
\label{energy}
\end{equation}
and the conserved angular momentum
\begin{equation}
j = -{\bf \eta \cdot u} = \Omega R^2 \sin^2\theta,
\end{equation}

where $u^t$ is the time component of the four velocity $u$ and $\Omega$ is the angular velocity of Keplerian orbits in Schwarzschild geometry. The condition ${\bf u \cdot u} = -1$ in case of the Schwarzschild metric gives
\begin{equation} 
u^t = \gamma_{jet} = \left(1-2 M/R-\frac{\dot{R}^2/c^2}{1-2 M/R}-R^2 \dot{\theta}^2/c^2-R^2 \sin^2 \theta \ \dot{\phi}^2/c^2\right)^{-1/2}.
\label{gammajet}
\end{equation}

This can be used to write the general expression for the velocity $\beta_{jet}$ of the emitting spot as 
\begin{align} \label{betajet}
\beta_{jet} &= \frac{\dot{R}^2/c^2}{1-2 M/R}-R^2 \dot{\theta}^2/c^2-R^2 \sin^2 \theta \ \dot{\phi}^2/c^2 \\ \nonumber
 &= \left(1-2 M/R-1/\gamma^2_{jet}\right)^{1/2}
\end{align}

\begin{figure}
\centerline{\includegraphics[scale=.6,angle=90]{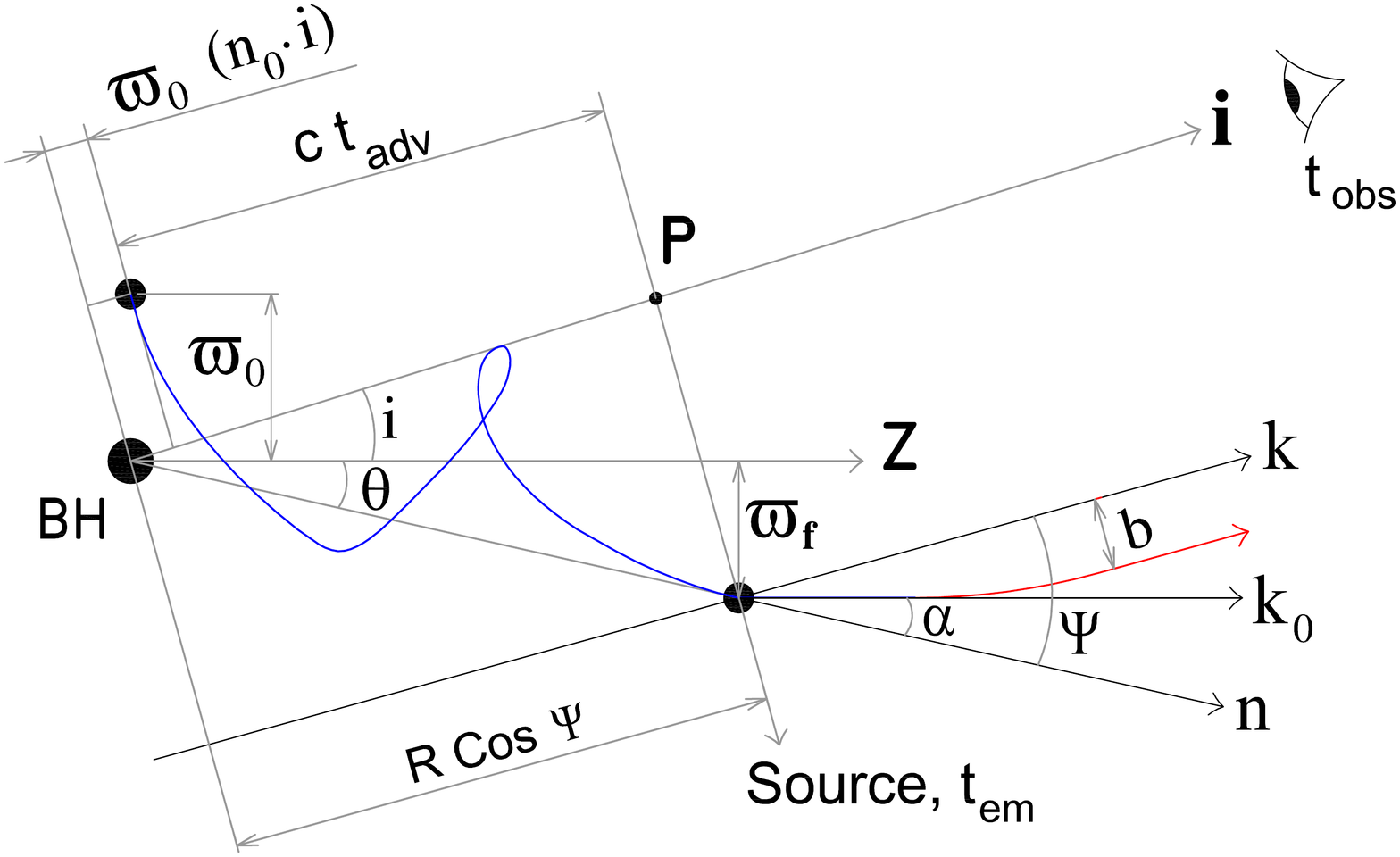}}
\caption{General helical flow geometry showing the path of the emitting source on the jet (blue) launched from the cylindrical radius $\varpi_0$ emitting a light ray which is subjected to the light bending effect (red). If the time of emission is $t_{em}$, the time at which the signal is observed $t_{obs} = \left(t_{em}-t_{adv} + \Delta t_{LB}\right) (1+z)$ where $t_{adv}$ is the time taken for the source to traverse the distance $R \cos \psi$ and $\Delta t_{LB}$ is the correction due to light bending effect in curved space-time. The blob trajectory is shown in blue, emission vectors in black and geometrical quantities including distances in grey. {\em See the online article for the color version.}}
\label{emmgeom}
\end{figure}

The various emission and direction vectors include the radial vector ${\bf n}$, the direction vector of the initially emitted light ray ${\bf k_0}$ and the final direction vector pointing along the observer line of sight ${\bf k}$. A general path showing the source motion along a helical trajectory and the emission geometry along with the above vectors is presented in a schematic in Fig. \ref{emmgeom}.

The vector ${\bf k_0}$ is given in terms of ${\bf k}$ and ${\bf n}$ as
\begin{equation}
{\bf k_0} = \frac{\sin \alpha}{\sin \psi} {\bf k}+\frac{\sin(\psi-\alpha)}{\sin \psi} {\bf n};
\label{veck0}
\end{equation}
this identity can be verified applying dot and cross products of $\bf k$ and $\bf n$ and using their geometry given in Fig. \ref{emmgeom}.

In the Cartesian $(x,y,z)$ coordinate system centred on the black hole, the components of the direction vector ${\bf k}$ are given by

\begin{equation}
{\bf k} = (\sin i,0,\cos i)
\label{veck}
\end{equation}
and the components of the emission vector ${\bf n}$ are given by
\begin{equation}
{\bf n} = (\sin \theta \cos \phi,\sin \theta \sin \phi,\cos \theta).
\label{vecn}
\end{equation}

The initial emission angle $\alpha$ can be written in terms of the initial direction vector ${\bf k_0}$ and the emission vector ${\bf n}$ as $\cos \alpha = {\bf k_0} \cdot {\bf n}$. The final angle of the emitted ray $\psi$ can be written in terms of the final direction vector ${\bf k}$ and the emission vector ${\bf n}$ as $\cos \psi = {\bf k} \cdot {\bf n}$. This can be expanded as 
\begin{equation}
\cos \psi = \cos i \cos \theta+\sin i \sin \theta \cos \phi.
\end{equation}

Assume that an emitter located at a radial distance $R$ emits a light ray at an angle $\alpha$ with respect to the radial vector. The light ray undergoes bending due to space-time curvature to emerge at a final angle $\psi$ with respect to the radial vector and the relationship between $\psi$ and $\alpha$ can be determined by solving and combining %(e.g. \citealt{2002ApJ...566L..85B,2006MNRAS.373..836P})
\begin{equation}
\psi = \int^{\infty}_{R} \frac{dr}{r^2} (1/b^2-1/r^2 (1-2 M/r))^{-1/2}
\end{equation}
and
\begin{equation}
\sin \alpha = \frac{b}{R} (1-2 M/R)^{1/2},
\label{sinalpha}
\end{equation}
where $b$ is the impact parameter shown in Fig. \ref{emmgeom}. The derivation of the above equation is presented in Appendix \ref{gfacderivation}. This propagation effect can be approximated by the expression \citep{2002ApJ...566L..85B}
\begin{equation}
1-\cos \alpha = (1-\cos \psi) (1-2 M/R).
\label{lightbending}
\end{equation}

Using the above approximation, we find 
\begin{equation}
\frac{\sin \alpha}{\sin \psi} = (1-2 M/R)^{1/2} (1-2 M/R+(4 M/R) (1+\cos \psi)^{-1})^{1/2}.
\label{sinapprox}
\end{equation}

The travel time will be different for photons emitted from different radial locations $R$. The difference between the travel time for a ray emitted from a position $R$ and that from a position at the centre of the coordinate system can be obtained from the integral
\begin{equation}
\Delta t =  \int^{\infty}_{R} \frac{dr}{1-2 M/r} \left(\left(1-\frac{b^2}{r^2} (1-2 M/r)\right)^{-1/2}-1\right).
\end{equation}

The time delay due to light bending can be approximated \citep{2006MNRAS.373..836P} as

\begin{equation}
\Delta t_{LB} = \left(\frac{\delta}{8} y^2 \left(1+y \left(\frac{1}{3}-\frac{2 \delta}{14}\right) \right) \right) \frac{R}{c},
\label{timedelaySch}
\end{equation}
where $y = (1-\cos \psi)$ and $\delta = 2 M/R$. The time delay is defined with respect to straight paths from the point $P$ to the observer as shown in Fig. \ref{emmgeom}.

The time of advance due to orbital motion of the emitting source, $t_{adv}$ can be expressed in terms of geometric factors from Fig. \ref{emmgeom} as
\begin{equation}
t_{adv} = (1-y) R/c - ({\bf n_0\cdot i}) \varpi_0/c,
\end{equation}
where the constant shift $({\bf n_0\cdot i})$ accounts for the initial offset in $R$ in the projected ${\bf i}$ direction towards the observer. If the time of emission in the source frame is $t_{em}$, the observed period of the signal $t_{obs}$ is reduced due to the effects of the component of the source moving towards the observer line of sight and due to the time delay induced by light bending and the disk inclination and is given by
\begin{align}\label{tobs1}
t_{obs} &= \left[t_{em}-t_{adv} + \Delta t_{LB}\right] (1+z) = \left[t_{em}+(y-1) R/c + ({\bf n_0\cdot i}) \varpi_0/c + \Delta t_{LB}\right] (1+z) \\ \nonumber
&= \left[t_{em}- (R/c) \cos \psi + ({\bf n_0\cdot i}) \varpi_0/c + \Delta t_{LB}\right] (1+z).
\end{align}
This is used to map the time of emission to the time of observation by accounting for light bending where $\Delta t_{LB}$ is the time delay from eqn. (\ref{timedelaySch}) due to light bending and where $z$ is the cosmological redshift of the emitting source. The effect of the time delay due to the light bending and disk inclination is to change the phase of reception of the observed signal. The phase change is expected to systematically increase as the emitting source moves along the expanding jet towards the observer. Also, we set the constant offset so that $t_{obs} (t_{em}=0) = 0$ for no light bending.

\section{Flow trajectories in and beyond Zone 3 and resulting light curves}
\label{kinmodels}

\begin{figure}
\includegraphics[scale=0.4]{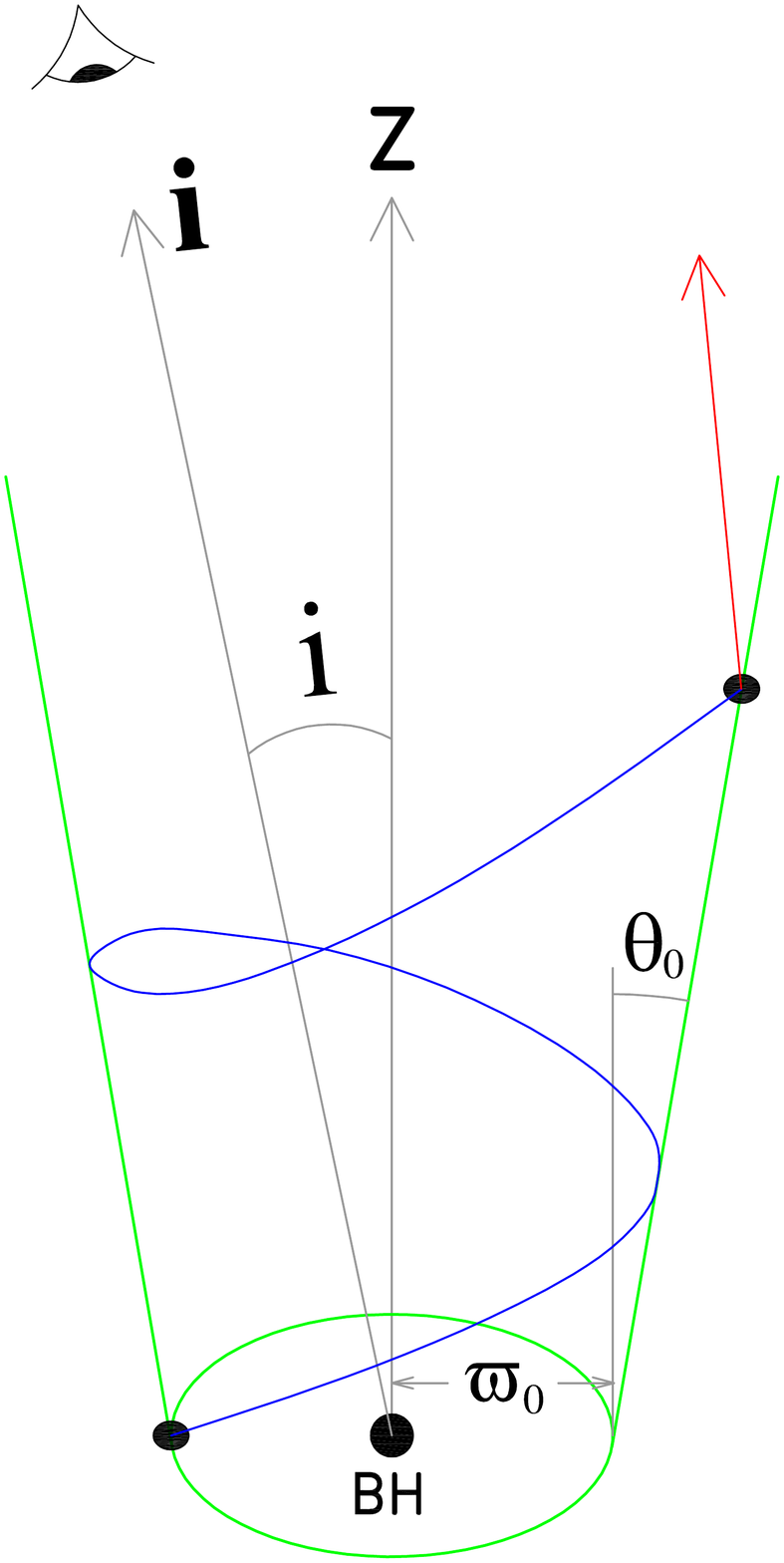} \includegraphics[scale=.4]{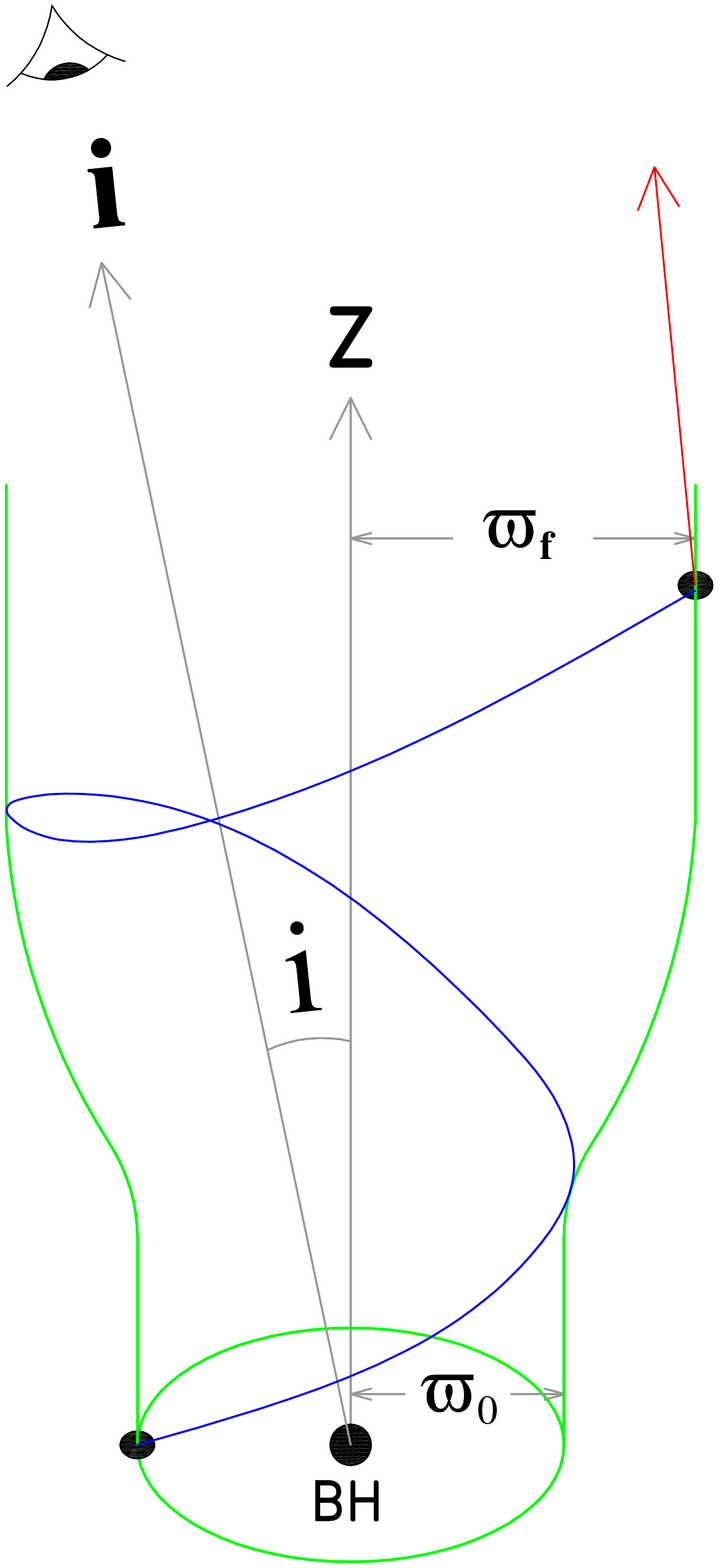}
\caption{Helical trajectory of an emitting blob in Schwarzschild geometry, constrained along rotating magnetic field lines with footpoints on a Keplerian disk (at cylindrical radius $\varpi_o$). In the cone model (left plot), the half opening angle of the jet is $\theta_0$. In the funnel model (right plot), the flow is asymptotically bound by a cylinder of radius $\varpi_f$ at large $z$. The blob trajectory is shown in blue, emission vector in red, geometrical quantities including distances in grey and the jet magnetic surface shape in green. {\em See the online article for the color version.}}
\label{funnelgeom}
\end{figure}

We now discuss the construction of the light curve for the variability model. The motion of the flux frozen blob is along a magnetic surface with foot points anchored on the accretion disk. A basic schematic of the blob motion along the magnetic surface is presented in Fig. \ref{funnelgeom}.

This surface co-rotates along with the disk at an angular frequency given by
\begin{equation}
\Omega_F = \frac{c M^{1/2}}{\varpi^{3/2}_F+a M^{1/2}},
\label{omegaf}
\end{equation}
where $a$ is the black hole spin parameter and $\varpi_F$ is the cylindrical radius to the foot point from the coordinate system centred on the black hole ($z = 0$). The radius of the light cylinder $\varpi_L$ is the position on a cylindrical surface along which the plasma moves at the speed of light and calculated as 
\begin{equation}
\varpi_L = \frac{c}{\Omega_F} = \frac{\varpi^{3/2}_F}{M^{1/2}}+a.
\label{rhoL}
\end{equation}
The cylindrical radius from which the blob is launched is given by $\varpi_0 = f \varpi_L$ where $f$ = 0.1 - 10. The specific angular momentum of the blob, $j$ is conserved along the magnetic surface. This occurs beyond the Alfv\'{e}n point with a cylindrical radius $\varpi_A = x_A \varpi_L$ where $x^2_A \leq 1$. The conserved angular momentum is then given by
\begin{equation}
j = j_{\infty} = x^2_A \varpi^2_L \Omega_F.
\end{equation}

The $\Omega_F$ in eqn. (\ref{omegaf}) is the angular frequency of orbits in the Kerr metric and we make use of $a \neq 0$ only for comparison with the special relativistic model presented in \cite{1992A&A...255...59C}. For the general relativistic models, we make use of the Schwarzschild geometry ($a = 0$) to maintain consistency with the treatment of the Doppler boost factor, aberration and light bending.

The instantaneous position of the emitting spot is given by
\begin{equation}
{\bf x_s} = (x,y,z) = (\varpi (t) \cos \phi (t), \varpi (t) \sin \phi (t), z(t)),
\end{equation}
where $\varpi$ is the cylindrical distance to the emitting spot from $z(t)$. In our variability model, a prescription for $\varpi = \varpi (z)$ is  given by the geometry of the magnetic surface. The velocity components (${\bf \dot{x}_s} = d{\bf x_s}/dt$) are then
\begin{equation}
{\bf \dot{x}_s} = (\dot{x},\dot{y},\dot{z}) = (\dot{\varpi} \cos \phi (t)-\varpi \dot{\phi} (t), \dot{\varpi} \sin \phi (t)+\varpi \dot{\phi} (t), \dot{z}).
\label{xsdot}
\end{equation}
The angle between the velocity vector of the spot ${\bf \dot{x}_s}$ and the initial direction vector ${\bf k_0}$ is given by
\begin{equation}
\cos \xi = \frac{{\bf \dot{x}_s} \cdot {\bf k_0}}{|{\bf \dot{x}_s}|} = \frac{1}{|{\bf \dot{x}_s}|} \left(\frac{\sin \alpha}{\sin \psi} {\bf \dot{x}_s} \cdot {\bf k}+\frac{\sin (\psi-\alpha)}{\sin \psi} {\bf \dot{x}_s} \cdot {\bf n}\right).
\end{equation} 

Using the initial direction vector ${\bf k_0}$ from eqn. (\ref{veck0}), the final direction vector ${\bf k}$ from eqn. (\ref{veck}) and the radial vector ${\bf n}$ from eqn. (\ref{vecn}), we can write the most general form of $\cos \xi$ as 
\begin{align} \label{cosxi}
\cos \xi &= \frac{1}{|{\bf \dot{x}_s}|} \left(\frac{\sin \alpha}{\sin \psi} {\bf \dot{x}_s} \cdot {\bf k}+ \frac{\sin(\psi-\alpha)}{\sin \psi} {\bf  \dot{x}_s} \cdot {\bf n}\right) \\ \nonumber
 &= \frac{1}{|{\bf  \dot{x}_s}|} \left \{ \frac{\sin \alpha}{\sin \psi}[\dot{\varpi} \cos \phi \sin i- \varpi \Omega \sin \phi \sin i+\dot{z} \cos i]+ \frac{\sin(\psi-\alpha)}{\sin \psi} \dot{z} [\tan \theta_0 \sin \theta+\cos \theta]\right\} \\ \nonumber
 &= \frac{1}{|{\bf  \dot{x}_s}|} \left \{ \dot{z} \left(\frac{\sin \alpha}{\sin \psi} [\tan \theta_0 \cos \phi \sin i+\cos i]+\frac{\sin(\psi-\alpha)}{\sin \psi} [\tan \theta_0 \sin \theta+\cos \theta] \right.\right. \\ \nonumber
&\left.\left.-\varpi \Omega \frac{\sin \alpha}{\sin \psi} \sin \phi \sin i\right) \right\}.
\end{align}

The Doppler factor $D$ evaluated in an instantaneous stationary frame at the source using the emission vectors given in eqns. (\ref{veck0}), (\ref{veck}) and (\ref{vecn}) is given by% \citep{2006MNRAS.373..836P},
\begin{equation}
D = \frac{1}{\gamma_{jet} (1-\beta_{jet} \cos \xi)}.
\end{equation}

In the observed frame, the above expression must be modified to account for the gravitational redshift. Thus, the effective redshift factor $g$ which is the ratio of observed to emitted energy of the emitted ray must include the above Doppler factor as well as the gravitational redshift effect and is given by
\begin{equation}
g = \frac{E_{\mathrm{observed}}}{E_{\mathrm{emitted}}} = (1-2 M/R)^{1/2} D = \frac{(1-2 M/R)^{1/2}}{\gamma_{jet} (1-\beta_{jet} \cos \xi)}.
\label{gfacdef}
\end{equation}

An alternate derivation of the above equation is presented in Appendix \ref{gfacderivation}. If $\gamma_{jet,i}$ is the bulk Lorentz factor of the flow at the initial launch radius $\varpi_0$, the energy conservation condition along the trajectory can be written as
\begin{equation}
\frac{u^t}{(1-2 M/R)} = \frac{\gamma_{jet}}{(1-2 M/R)} = \varepsilon = \frac{\gamma_{jet,i}}{(1-2 M/\varpi_0)} = \gamma_{jet,f}.
\label{energyin}
\end{equation}

This can be used to express $\gamma_{jet}$ in terms of the initial parameters $\varpi_0$ and $\gamma_{jet,i}$. From the simulations invoking radiation pressure and drag in \S \ref{radprdynamics}, we motivate stable values of $\gamma_{jet,i} \sim 2 - 6$ at larger $x$ such as the initial launching point (foot points on the disk) which we have taken to be between $(0.1-10)$ light cylinder radii. The light curve is given by the spectral flux density observed $F_{\nu}(t)$. If $F^{'}_{\nu}$ is the spectral flux density in the co-moving frame, these are related by the expression
\begin{equation}
F_\nu(t) = g^{\lambda}(t) F^{'}_\nu(t),
\end{equation}

where $\lambda = 3+\Gamma$ for a resolved blob of plasma and $\lambda = 2+\Gamma$ for a continuous flow; $\Gamma$ is the spectral index which is the slope in the relation $F_\nu \propto \nu^{\Gamma}$ between the spectral flux $F_\nu$ and the emission frequency in the observer frame. Either case of $\lambda$ is possible depending on the particular application to observations, but for the purposes of illustration, we take $\lambda = 2 + \Gamma$ in order to compare our results with \cite{1992A&A...255...59C}.

As $t$ is the time in the source frame, we obtain the time in the observer frame using the transformation in eqn. (\ref{tobs1}) with the time delay $\Delta t_{LB}$ to obtain $F_\nu = F_\nu (t_{obs})$ in the observer frame. The beaming effect will be prominent in the light curve when the angle $\xi$ between the initial emission vector and the velocity vector of the source is close to the disk inclination angle $i$ between the normal drawn in the coordinate system centred on the black hole and the observer direction. The quasi-periodic behaviour from orbital signatures when beaming occurs is expected to be distinctly visible in observed light curve and could last for only a few cycles.

\subsection{Special relativistic cone trajectory}
\label{SRcone}

We construct a special relativistic jet model based on that presented in \cite{1992A&A...255...59C}. The model consists of a relativistic blob in a cone geometry as shown in Fig. \ref{funnelgeom}. The kinematical prescription includes expressions for the cylindrical radius $\varpi$ and the associated velocity $\dot{\varpi}$ in terms of $z$ and $\dot{z}$. In the cone model with a jet half-opening angle $\theta_0$,
\begin{equation}
\varpi = \varpi_0+z \tan \theta_0.
\label{rhozcone}
\end{equation}

Using the above equation, the velocity associated with $\varpi(t)$ is given by
\begin{equation}
\dot{\varpi} = \dot{z} \tan \theta_0.
\label{rhodotzdot}
\end{equation}

The conservation of angular momentum gives an azimuthal velocity $\dot{\phi} = \Omega(t) = j_\infty/\varpi^2$ and the phase is given by $\phi(t) = \int^t_0 \Omega (\tilde{t}) d\tilde{t}$. The condition ${\bf u \cdot u} = -1$ in case of the Minkowskian metric, an approximation used in \cite{1992A&A...255...59C} then gives
\begin{equation}
(u^t)^2 = \gamma^2_{jet} = \frac{1+u^2_p}{1-\varpi^2 \Omega^2/c^2} = \frac{1+u^2_p}{\disp 1-\frac{j^2_{\infty}}{\varpi^2 c^2}},
\label{gammajetCK}
\end{equation}
where $u_p = u^t (\dot{\varpi}^2+\dot{z}^2)^{1/2}$ is the poloidal velocity. In the region where collimation of the jet occurs ($\varpi >> \varpi_L$), the bulk Lorentz factor $\gamma_{jet}$ becomes a constant due to energy conservation as eqn. (\ref{energy}) can be written for large $R$ as $\gamma_{jet}$ = $\varepsilon$. At large $R$, $\varpi$ is also large. Hence, $u_p$ tends to a constant value and like \citep{1992A&A...255...59C}, we make this approximation. For a jet half-opening angle $\theta_0$, as $\dot{z} = c u_p \cos \theta_0/\gamma_{jet}$, we obtain
\begin{equation}
\dot{z} = \frac{u_p}{\sqrt{1+u^2_p}} \cos \theta_0 \left(c^2-\frac{j^2_{\infty}}{\varpi^2}\right)^{1/2}.
\label{ztCK}
\end{equation}

The cylindrical distance $\varpi = \varpi(t)$ in this particular case due to the simplification of $u_p$ being a constant. From eqn. (\ref{rhodotzdot}), $\dot{\varpi} = \dot{z} \tan \theta_0$. Using $\dot{z}$ from eqn. (\ref{ztCK}) in this,
\begin{equation}
\dot{\varpi} =  \frac{c u_p}{\sqrt{1+u^2_p}} \sin \theta_0 \left(c^2-\frac{j^2_{\infty}}{\varpi^2}\right)^{1/2}.
\end{equation}
The above expression can be written in terms of $\dot{\varpi}$ and $\varpi$ as
\begin{equation}
\frac{\varpi \dot{\varpi}}{\disp\sqrt{\varpi^2-\frac{j^2_{\infty}}{c^2}}} = \frac{c u_p \sin \theta_0}{\sqrt{1+u^2_p}}.
\end{equation}
On integrating the above equation, we obtain 
\begin{equation}
\varpi(t) = \frac{j_{\infty}}{c} \left(1+\left(\sqrt{\frac{\varpi^2_0 c^2}{j^2_{\infty}}-1}+\frac{c^2 u_p}{j_{\infty} \sqrt{u^2_p+1}} \sin \theta_0 t \right)^2\right)^{1/2}.
\label{rhotCK}
\end{equation}

The magnitude of the source velocity is given by
\begin{equation}
|{\bf \dot{x}_s}| = ({\bf x_s}\cdot{\bf x_s})^{1/2} = (\dot{\varpi}^2+\varpi^2 \Omega^2+\dot{z}^2)^{1/2} = (\dot{z}^2 \sec^2 \theta_0+j^2_{\infty}/\varpi^2)^{1/2}
\end{equation}
In the absence of the light bending effect, $\psi = \alpha$ in which case the direction vectors $k_0 = k$. The angle between the velocity vector ${\bf x_s}$ and the initial direction vector ${\bf k_0}$, $\cos \xi$ from eqn. (\ref{cosxi}) then reduces to
\begin{equation}
\cos \xi = \frac{1}{|{\bf \dot{x}_s}|} (\dot{z} (\tan \theta_0 \cos \phi \sin i+\cos i) - \varpi \Omega \sin \phi \sin i).
\label{cosxiCK}
\end{equation}
The velocity $\dot{z}$ can be obtained from eqn. (\ref{ztCK}) and $\varpi$ from eqn. (\ref{rhotCK}). The effective redshift factor $g(t)$ is then given by 
\begin{equation}
g = \frac{1}{\gamma_{jet} (1-\beta_{jet} \cos \xi)}.
\end{equation}
We can use $\cos \xi$ from eqn. (\ref{cosxiCK}), $\gamma_{jet}$ from eqn. (\ref{gammajetCK}), $\beta_{jet} = \sqrt{1-1/\gamma^2_{jet}}$ and $\dot{\varpi}$ from eqn. (\ref{rhotCK}) in the above equation to obtain $g = g(t)$. The light curve is given by $F(t) = g^{2+\Gamma}(t)$. As $t$ is the time in the source frame, we obtain the time in the observer frame using the transformation in eqn. (\ref{tobs1}) without the time delay $\Delta t_{LB}$ in this case to obtain $F = F(t_{obs})$ in the observer frame. 

\begin{figure}[!h]
\centerline{\includegraphics[scale=0.35]{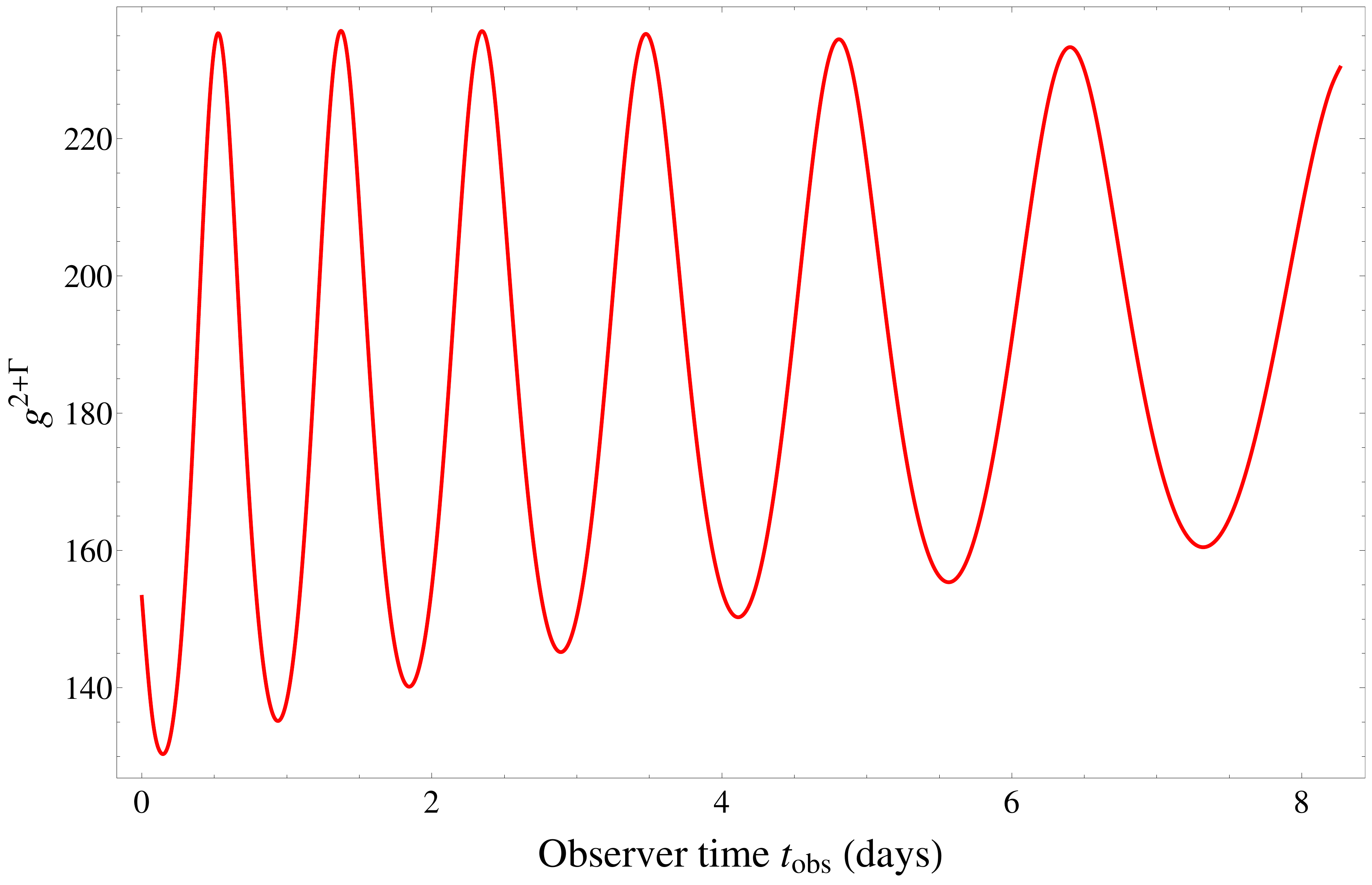}}
\caption{Simulated light curve for $u_p = 3$, $x_A = 0.9$, $i = 0.05^{\circ}$, $M_{\bullet} = 5 \times 10^7 M_{\odot}$, $a = 0.8$, $\varpi_0 = 10 \varpi_L$ and $\Gamma = 1$. The light curve (red curve) shows the quasi-periodic oscillation expected from a general BL Lac object when the special relativistic formalism is used in the simulation. {\em See the online article for the color version.}}
\label{CK92}
\end{figure}

To illustrate the validity of the model developed in \S \ref{tframework}, we simulate a light curve for a general BL Lac object using the expressions presented in this section. For this, we consider the same parameters which were considered in \cite{1992A&A...255...59C}, i.e. $u_p = 3$, $i = 0.05^{\circ}$, $M_{\bullet} = 5 \times 10^7 M_{\odot}$, $a = 0.8$ and $\varpi_0 = 10 \varpi_L$. The parameter $u_p = 3$ is chosen in order that $\gamma_{jet,i}$ at large $\varpi \sim$ 3.2 which is within the range predicted by the effects of radiation pressure and drag in \S \ref{radprdynamics} and is in the same range as that considered in \cite{1992A&A...255...59C}. The resulting light curve is plotted in Fig. \ref{CK92} which is similar to Fig 3 of \cite{1992A&A...255...59C} with phase shift.

\subsection{Fully relativistic cone model}
\label{GRcone}

We construct a fully relativistic jet model in Schwarzschild geometry which consists of a relativistic blob in a cone geometry as shown in Fig. \ref{funnelgeom}. Here we consider general relativistic effects which include time delay due to light bending, the Doppler and gravitational redshift and the treatment of the poloidal velocity as a general function of the geometrical and kinematical parameters as opposed to the constant assumed in the previous case. The kinematical prescription includes expressions for the cylindrical radius $\varpi$ and the associated velocity $\dot{\varpi}$ in terms of $z$ and $\dot{z}$. 

The cylindrical distance $\varpi(z)$ is given by eqn. (\ref{rhozcone}) and the velocity associated with $\varpi (z)$ is given by eqn. (\ref{rhodotzdot}). The radial distance $R(z)$ is given by 
\begin{equation}
R(z) = (\varpi^2(z)+z^2)^{1/2} = (\varpi^2_0+z^2 \sec^2 \theta_0+2 \varpi_0 z \tan \theta_0)^{1/2}
\label{Rzcone}
\end{equation}
and the velocity associated with $R(z)$ is given by
\begin{equation}
\dot{R} = \frac{1}{R} (\varpi \dot{\varpi}+z \dot{z}) = \frac{\dot{z}}{R} (\varpi_0 \tan \theta_0+z \sec^2 \theta_0).
\end{equation}

The polar angle $\theta(z)$ is given by
\begin{equation}
\sin \theta (z) = \varpi(z)/R(z)
\end{equation}
and the velocity associated with $\theta(z)$ is given by
\begin{equation}
R \dot{\theta} = \frac{1}{z} \left(\dot{\varpi}-\frac{\varpi \dot{R}}{R}\right) = -\frac{\dot{z} \varpi_0}{R}.
\end{equation}

The conservation of angular momentum gives an azimuthal velocity $\dot{\phi} = \Omega(z) = j_\infty/\varpi^2$. Using the condition ${\bf u \cdot u} = -1$, from eqn. (\ref{gammajet})
\begin{equation} 
u^t = \gamma_{jet} = \left((1-2 M/R)-\frac{\dot{z}^2}{R^2 c^2} \left(\varpi^2_0+\frac{(\varpi_0 \tan \theta_0+z \sec^2 \theta_0)^2}{(1-2 M/R)}\right)-\frac{j^2_\infty}{\varpi^2 c^2}\right)^{-1/2}.
\label{gammajetcone}
\end{equation}

From the energy conservation condition along the trajectory,
\begin{equation}
\frac{u^t}{(1-2 M/R)} = \varepsilon = \frac{\gamma_{jet,i}}{(1-2 M/\varpi_0)},
\end{equation}
where $\gamma_{jet,i}$ is the initial Lorentz factor. Using $\gamma_{jet,i}$ from eqn. (\ref{gammajetCK}) for the case of $\varpi = \varpi_0$ at $z=0$ and a constant $u_p$, we obtain $\gamma_{jet,i} = \left(\frac{\disp 1+u^2_p}{\disp 1-\frac{j^2_{\infty}}{\varpi^2_0 c^2}}\right)^{1/2}$ and
\begin{equation}
u^t = \gamma_{jet} = \gamma_{jet,i} \frac{(1-2 M/R)}{(1-2 M/\varpi_0)}.
\end{equation}

Using the above expression for $u^t$ in eqn. (\ref{gammajetcone}), we obtain an equation for $\dot{z}$ given by
\begin{equation}
\dot{z} = c R \left((1-2 M/R)-\frac{j^2_\infty}{\varpi^2 c^2}-\frac{(1-2 M/\varpi_0)^2}{(1-2 M/R)^2} \frac{1}{\gamma^2_{jet,i}}\right)^{1/2} \left(\varpi^2_0+\frac{(\varpi_0 \tan \theta_0+z \sec^2 \theta_0)^2}{(1-2 M/R)}\right)^{-1/2}.
\label{dotzcone}
\end{equation}

In the above equation, $\varpi = \varpi(z)$ from eqn. (\ref{rhozcone}) and $R = R(z)$ from eqn. (\ref{Rzcone}). The above expression can be integrated to obtain $z = z(t)$ as
\begin{equation}
\int^z_0 \frac{dz}{R} \left(\frac{\disp \varpi^2_0+\frac{(\varpi_0 \tan \theta_0+z \sec^2 \theta_0)^2}{(1-2 M/R)}}{\disp(1-2 M/R)-\frac{j^2_\infty}{\varpi^2 c^2}-\frac{(1-2 M/\varpi_0)^2}{(1-2 M/R)^2} \frac{1}{\gamma^2_{jet,i}}}\right)^{1/2} = c \int^t_0 dt.
\end{equation}

Once we obtain $z = z(t)$, this can be used to evaluate $\dot{z} = \dot{z}(z(t))$, $\varpi = \varpi(z(t))$, $\dot{\varpi} = \dot{\varpi}(z(t))$. The magnitude of the source velocity is given by
\begin{equation}
|{\bf \dot{x}_s}| = ({\bf x_s}\cdot{\bf x_s})^{1/2} = (\dot{\varpi}^2+\varpi^2 \Omega^2+\dot{z}^2)^{1/2} = (\dot{z}^2 \sec^2 \theta_0+j^2_{\infty}/\varpi^2)^{1/2}.
\end{equation}

The velocity $\dot{z}$ = $\dot{z}(z(t))$ and all other terms in the above expression can be cast in terms of $z(t)$. In the current model, we consider the light bending effect in which case the direction vector ${\bf k_0}$ is obtained from eqn. (\ref{veck0}) and the angle between the velocity vector ${\bf x_s}$ and the initial direction vector ${\bf k_0}$ given by $\cos \xi$ is given by the general expression in eqn. (\ref{cosxi}).

\begin{figure}[!h]
\centerline{\includegraphics[scale=0.27]{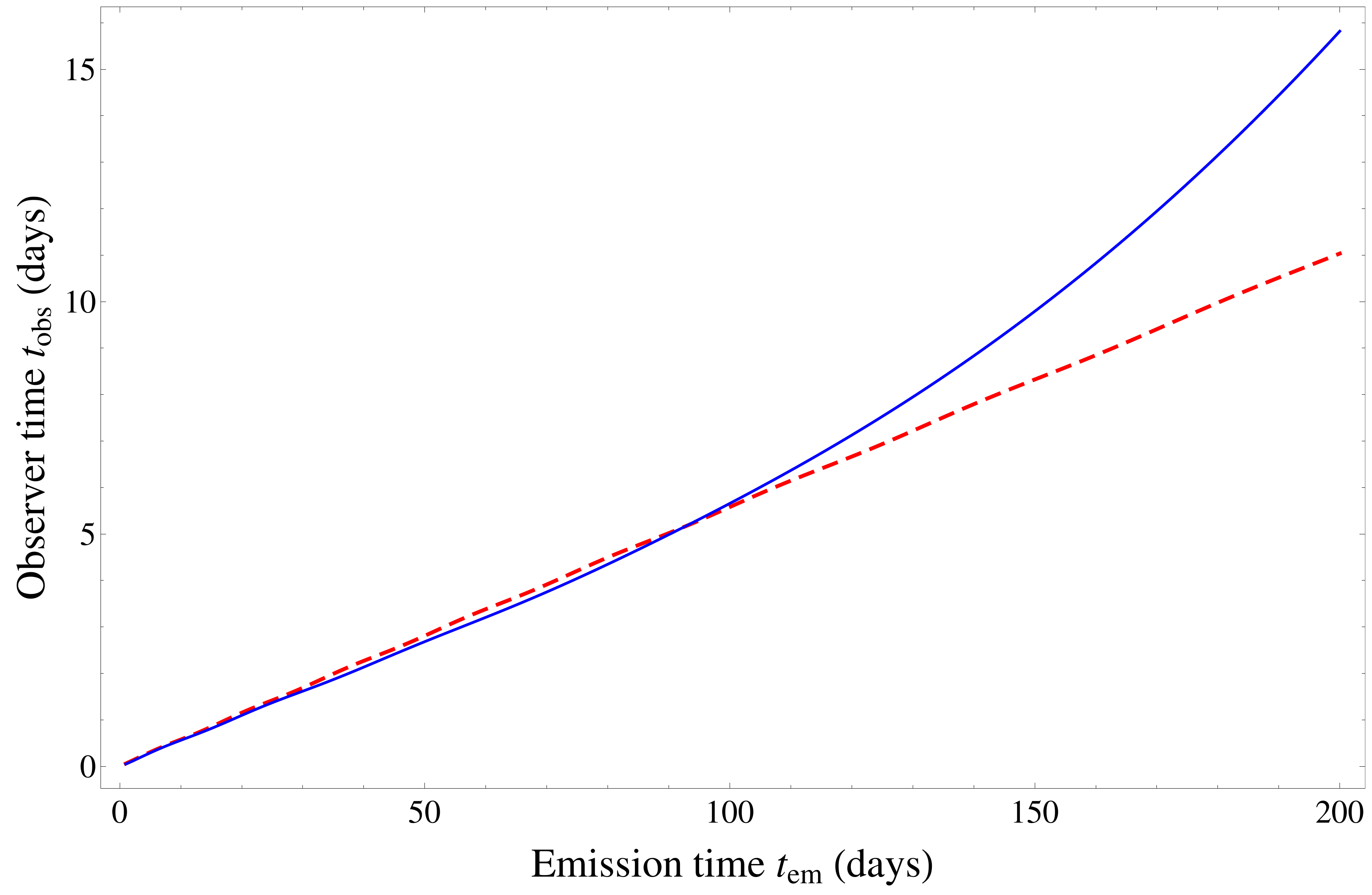}}
\caption{Curves showing $t_{obs}$ versus $t_{em}$ for the fully relativistic cone model (blue) and the special relativistic cone model (red, dashed). The $t_{obs}$ in the former increases systematically and reaches a difference of $\sim$ 5 days by the end of the simulation with respect to the latter. {\em See the online article for the color version.}}
\label{tplot1c}
\end{figure}

The effective redshift factor $g(t)$ is then given by the general expression in eqn. (\ref{gfacdef}). We can use $\cos \xi$ from eqn. (\ref{cosxi}), $\gamma_{jet}$ from eqn. (\ref{gammajetcone}), $\beta_{jet}$ from eqn. (\ref{betajet}) in the equation for $g$ to obtain $g = g(t)$. The light curve is given by $F(t_{obs}) = g^{2+\Gamma}(t_{obs})$, expressing $t$ in terms of $t_{obs}$. The importance of making use of a fully relativistic model when compared to the earlier special relativistic model is presented in terms of the difference between the $t_{obs}$ in both cases. It is seen that for the fully relativistic model, $t_{obs}$ tends to increase systematically with $t_{em}$. This difference in timescale is of the order of $\sim$ 5 days by the end of the simulation. This difference is plotted in Fig. \ref{tplot1c}, which justifies the use of the fully relativistic model.

To illustrate the model developed in \S \ref{tframework} and the comparison with the special relativistic cone case in \S \ref{SRcone}, we simulate a light curve for a general BL Lac object using the expressions presented in this section. Using again $u_p = 3$ as an initial condition, with $i = 0.05^{\circ}$, $M_{\bullet} = 5 \times 10^7 M_{\odot}$, $a = 0.8$ and $\varpi_0 = 10 \varpi_L$, the resulting light curve is shown in Fig. \ref{GRcone}. The parameter $u_p = 3$ is chosen in order that $\gamma_{jet,i}$ at large $\varpi \sim$ 3.2 which is within the range predicted by the effects of radiation pressure and drag in \S \ref{radprdynamics} and is in the same range as that considered in \cite{1992A&A...255...59C}. This is a comparison between the light curves in the fully relativistic cone case and the special relativistic case which indicates an increased amplitude in the former; there is also a systematically increasing phase lag compared to the latter case as expected due to the effect of time delay caused by light bending. The ratio of the amplitudes gives a 12 \% increase in the maximum amplitude due to the GR boost factor in $g$. The beaming effect is observed for the last two cycles for the general relativistic cone case where there is an increased amplitude which then settles down. The increased amplitude ratio and the phase lag change can be attributed to the integrals over time of intensity and phase amplifying the small but early differences in phase.  Even a small phase difference between the general relativistic and special relativistic simulations gets amplified due to the nature of the $g$ factor in eqn. (\ref{gfacdef}) which includes the gravitational redshift factor $(1-2 M/R)^{1/2}$ and light bending effects.

\begin{figure}[!h]
\centerline{\includegraphics[scale=0.27]{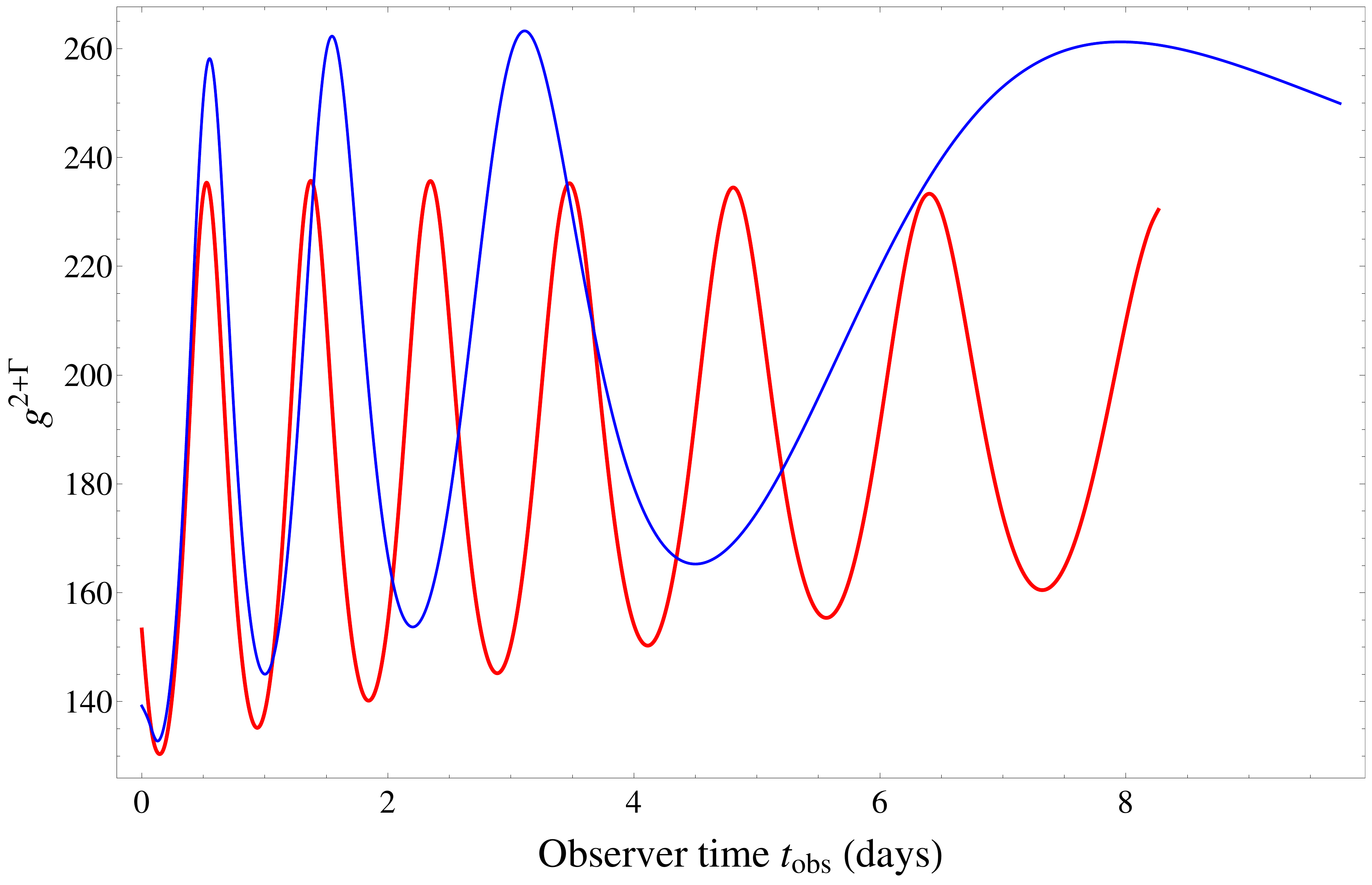}}
\caption{Simulated light curve for $u_p = 3$, $x_A = 0.9$, $i = 0.05^{\circ}$, $M_{\bullet} = 5 \times 10^7 M_{\odot}$, $a = 0.8$, $\varpi_0 = 10 \varpi_L$ and $\Gamma = 1$. The light curve shows the quasi-periodic oscillation expected from a general BL Lac object. A general relativistic light curve from orbital features in a cone geometry (blue curve) is compared with the previously calculated special relativistic light curve from orbital features in a cone geometry (red curve). The increased amplitude ratio and the phase lag change can be attributed to the integrals over time of intensity and phase amplifying the small but early differences in phase. Even a small phase difference between the general relativistic and special relativistic simulations gets amplified due to the nature of the $g$ factor in eqn. (\ref{gfacdef}) which includes the gravitational redshift factor $(1-2 M/R)^{1/2}$ and light bending effects. {\em See the online article for the color version.}}
\label{GRcone}
\end{figure}

\subsection{Fully relativistic funnel model}
\label{GRfunnel}

We construct a general relativistic (Schwarzschild geometry) jet model which consists of a relativistic blob in a funnel geometry as shown in Fig. \ref{funnelgeom}. The shape of the magnetic surface can be determined by solving the relativistic Grad-Shafranov equation (e.g. \citealt{2001A&A...365..631F}). This is expected to yield stable, axi-symmetric magnetic field configurations. Here, we make an approximation of this shape and impose the condition of angular momentum conservation along this surface. Here too, we consider the general relativistic effects which include the time delay due to light bending, the Doppler and gravitational redshift and the treatment of the poloidal velocity as a general function of the geometrical and kinematical parameters as opposed to a constant that is assumed in the special relativistic case. The kinematical prescription includes expressions for the cylindrical radius $\varpi$ and the associated velocity $\dot{\varpi}$ in terms of $z$ and $\dot{z}$. The cylindrical distance $\varpi(z)$ is given by
\begin{equation}
\varpi(z) = \varpi_0 (1+k(1-e^{-z/z_f})),
\label{rhozfunnel}
\end{equation}
where $k = (\varpi_f-\varpi_0)/\varpi_0$, $\varpi_f$ is the cylindrical distance between the normal axis and the source position at the location where the funnel transitions into a cylinder. If we use $\varpi_f = q \varpi_L$, 
\begin{equation}
k_f = (\varpi_f-\varpi_0)/\varpi_0 = q/f - 1.
\label{kf}
\end{equation}
A constraint on $z_f$ can also be obtained based on the vertical distance $z$ at the region where $\varpi = \varpi_{f}$. If the jet half opening angle is $\theta_0$, $\tan \theta_0 = (\varpi_f-\varpi_0)/z_f$. Then 
\begin{equation}
z_f = (q-f) \varpi_L/\tan \theta_0.
\label{zf}
\end{equation}
The velocity associated with $\varpi (z)$ is given by
\begin{equation}
\dot{\varpi} = \dot{z} \frac{\varpi_0 k}{z_f} e^{-z/z_f}
\end{equation}

The radial distance $R(z)$ is given by 
\begin{equation}
R(z) = (\varpi^2(z)+z^2)^{1/2} = (\varpi^2_0 (1+k(1-e^{-z/z_f})^2+z^2)^{1/2}
\label{Rzfunnel}
\end{equation}
and the velocity associated with $R(z)$ is given by
\begin{equation}
\dot{R} = \frac{1}{R} (\varpi \dot{\varpi}+z \dot{z}) = \frac{\dot{z}}{R} \left(\frac{\varpi^2_0 k}{z_f} (1+k(1-e^{-z/z_f}) e^{-z/z_f}+z\right).
\end{equation}

As before, the polar angle $\theta(z)$ is given by
\begin{equation}
\sin \theta (z) = \varpi(z)/R(z)
\end{equation}
and the velocity associated with $\theta(z)$ is given by
\begin{equation}
R \dot{\theta} = \frac{1}{z} \left(\dot{\varpi}-\frac{\varpi \dot{R}}{R}\right) = \frac{\dot{z} \varpi_0}{R} (k e^{-z/z_f} (z/z_f+1)-(1+k)).
\end{equation}

The conservation of angular momentum gives an azimuthal velocity $\dot{\phi} = \Omega(z) = j_\infty/\varpi^2$. The condition ${\bf u \cdot u} = -1$ in case of the Schwarzschild metric gives
\begin{align} \label{gammajetfunnel}
u^t &= \left[(1-2 M/R) -\frac{j^2_\infty}{\varpi^2 c^2} \right.\\ \nonumber
&\left.-\frac{\dot{z}^2}{R^2 c^2} \left\{\left(\frac{\varpi^2_0 k}{z_f} (1+k(1-e^{-z/z_f}) ) e^{-z/z_f}+z\right)^2+\varpi^2_0 (k e^{-z/z_f} (z/z_f+1)-(1+k))^2\right\}\right]^{-1/2}.
\end{align}

%\begin{equation} \label{gammajetfunnel}
%u^t = \left((1-2 M/R) -\frac{j^2_\infty}{\varpi^2 c^2} -\frac{\dot{z}^2}{R^2 c^2} \left(\left(\frac{\varpi^2_0 k}{z_f} (1+k(1-e^{-z/z_f}) e^{-z/z_f}+z\right)^2+\varpi^2_0 (k e^{-z/z_f} (z/z_f+1)-(1+k))^2\right)\right)^{-1/2}.
%\end{equation}

Using the general expression for $u^t$ from eqn. (\ref{gammajetcone}), we obtain an equation for $\dot{z}$ given by
\begin{align}
\dot{z} &= c R \left((1-2 M/R)-\frac{j^2_\infty}{\varpi^2 c^2}-\frac{(1-2 M/\varpi_0)^2}{(1-2 M/R)^2} \frac{1}{\gamma^2_{jet,i}}\right)^{1/2} \\ \nonumber 
&\left[\left(\frac{\varpi^2_0 k}{z_f} (1+k(1-e^{-z/z_f})) e^{-z/z_f}+z\right)^2+\varpi^2_0 (k e^{-z/z_f} (z/z_f+1)-(1+k))^2\right]^{-1/2}.
\label{\partial otzfunnel}
\end{align}

In the above equation, $\varpi = \varpi(z)$ from eqn. (\ref{rhozfunnel}) and $R = R(z)$ from eqn. (\ref{Rzfunnel}). The above expression can be integrated to obtain $z = z(t)$ as
\begin{equation}
\int^z_0 \left(\frac{dz}{R} \frac{\disp \left(\frac{\varpi^2_0 k}{z_f} (1+k(1-e^{-z/z_f})) e^{-z/z_f}+z\right)^2+\varpi^2_0 (k e^{-z/z_f} (z/z_f+1)-(1+k))^2}{\disp(1-2 M/R)-\frac{j^2_\infty}{\varpi^2 c^2}-\frac{(1-2 M/\varpi_0)^2}{(1-2 M/R)^2} \frac{1}{\gamma^2_{jet,i}}}\right)^{1/2} = c \int^t_0 dt.
\end{equation}

If we use $k$ from eqn. (\ref{kf}) and $z_f$ from eqn. (\ref{zf}), we can obtain $z = z(t)$ which can then be used to evaluate $\dot{z} = \dot{z}(z(t))$, $\varpi = \varpi(z(t))$, $\dot{\varpi} = \dot{\varpi}(z(t))$. The magnitude of the source velocity is given by
\begin{equation}
|{\bf \dot{x}_s}| = ({\bf x_s}\cdot{\bf x_s})^{1/2} = (\dot{\varpi}^2+\varpi^2 \Omega^2+\dot{z}^2)^{1/2} = \left(\dot{z}^2 \left(\frac{\varpi^2_0 k^2}{z^2} e^{-2 z/z_f}+1\right) +j^2_{\infty}/\varpi^2\right)^{1/2}.
\end{equation}

Once we obtain $z = z(t)$, this can be used to evaluate $\dot{z} = \dot{z}(z(t))$, $\varpi = \varpi(z(t))$, $\dot{\varpi} = \dot{\varpi}(z(t))$. The velocity $\dot{z}$ = $\dot{z}(z(t))$ and all other terms in the above expression can be cast in terms of $z(t)$. In the current model, we consider the light bending effect in which case the direction vector ${\bf k_0}$ is obtained from eqn. (\ref{veck0}) and the angle between the velocity vector ${\bf x_s}$ and the initial direction vector ${\bf k_0}$ given by $\cos \xi$ is given by the general expression in eqn. (\ref{cosxi}).

\begin{figure}[!h]
\centerline{\includegraphics[scale=.27]{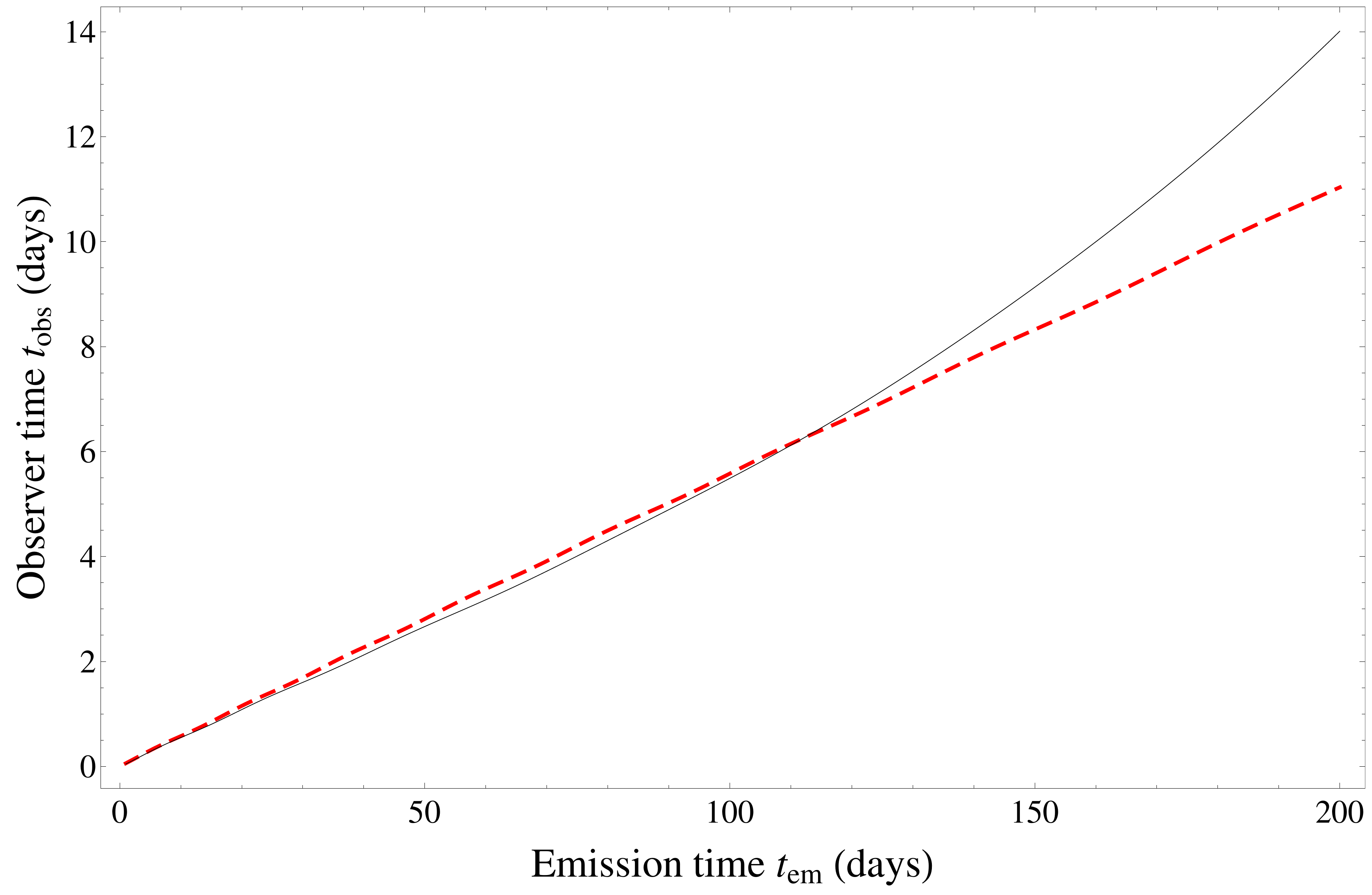}}
\caption{Curves showing $t_{obs}$ versus $t_{em}$ for the fully relativistic funnel model (black) and the special relativistic cone model (red, dashed). The $t_{obs}$ in the former increases systematically and reaches a difference of $\sim$ 3 days by the end of the simulation with respect to the latter. {\em See the online article for the color version.}}
\label{tplot1f}
\end{figure}

The effective redshift factor $g(t)$ is then given by the general expression in eqn. (\ref{gfacdef}). We can use $\cos \xi$ from eqn. (\ref{cosxi}), $\gamma_{jet}$ from eqn. (\ref{gammajetcone}),  $\beta_{jet}$ from eqn. (\ref{betajet}) in the equation for $g$ to obtain $g = g(t)$. The light curve is given by $F(t_{obs}) = g^{2+\Gamma}(t_{obs})$, expressing $t$ in terms of $t_{obs}$. In this case too, $t_{obs}$ tends to increase systematically with $t_{em}$ analogous to the full relativistic cone case. Though, the change is slightly gradual. This difference in timescale is of the order of $\sim$ 3 days by the end of the simulation. This difference is plotted in Fig. \ref{tplot1f}; clearly the use of the fully relativistic model impacts the phase significantly.

To illustrate the model developed in \S \ref{tframework} and the comparison with the special relativistic cone case in \S \ref{SRcone}, we simulate a light curve for a general BL Lac object using the expressions presented in this section. Using once again $u_p = 3$ as an initial condition, $i = 0.05^{\circ}$, $M_{\bullet} = 5 \times 10^7 M_{\odot}$, $a = 0.8$ and $\varpi_0 = 10 \varpi_L$, the light curve is plotted in Fig. \ref{GRfunnel}. The parameter $u_p = 3$ is chosen in order that $\gamma_{jet,i}$ at large $\varpi \sim$ 3.2 which is within the range predicted by the effects of radiation pressure and drag in \S \ref{radprdynamics} and is in the same range as that considered in \cite{1992A&A...255...59C}. A comparison between the light curves in the general relativistic cone case and the special relativistic case indicates an increased amplitude in the former in the funnel geometry as well; there is also a systematically increasing phase lag compared to the latter case as expected due to the effect of time delay caused by light bending. The ratio of the amplitudes gives a 12 \% increase in the maximum amplitude due to the GR boost factor in $g$ justifying the use of a fully relativistic model. Thus, the beaming effect is observed for the first two cycles for the general relativistic funnel case where there is an increased amplitude which then settles down. The increased amplitude ratio and the phase lag change here too can be attributed to the integrals over time of intensity and phase amplifying the small but early differences in phase.  Even a small phase difference between the general relativistic and special relativistic simulations gets amplified due to the nature of the $g$ factor in eqn. (\ref{gfacdef}) which includes the gravitational redshift factor $(1-2 M/R)^{1/2}$ and light bending effects.

\begin{figure}[!h]
\centerline{\includegraphics[scale=0.27]{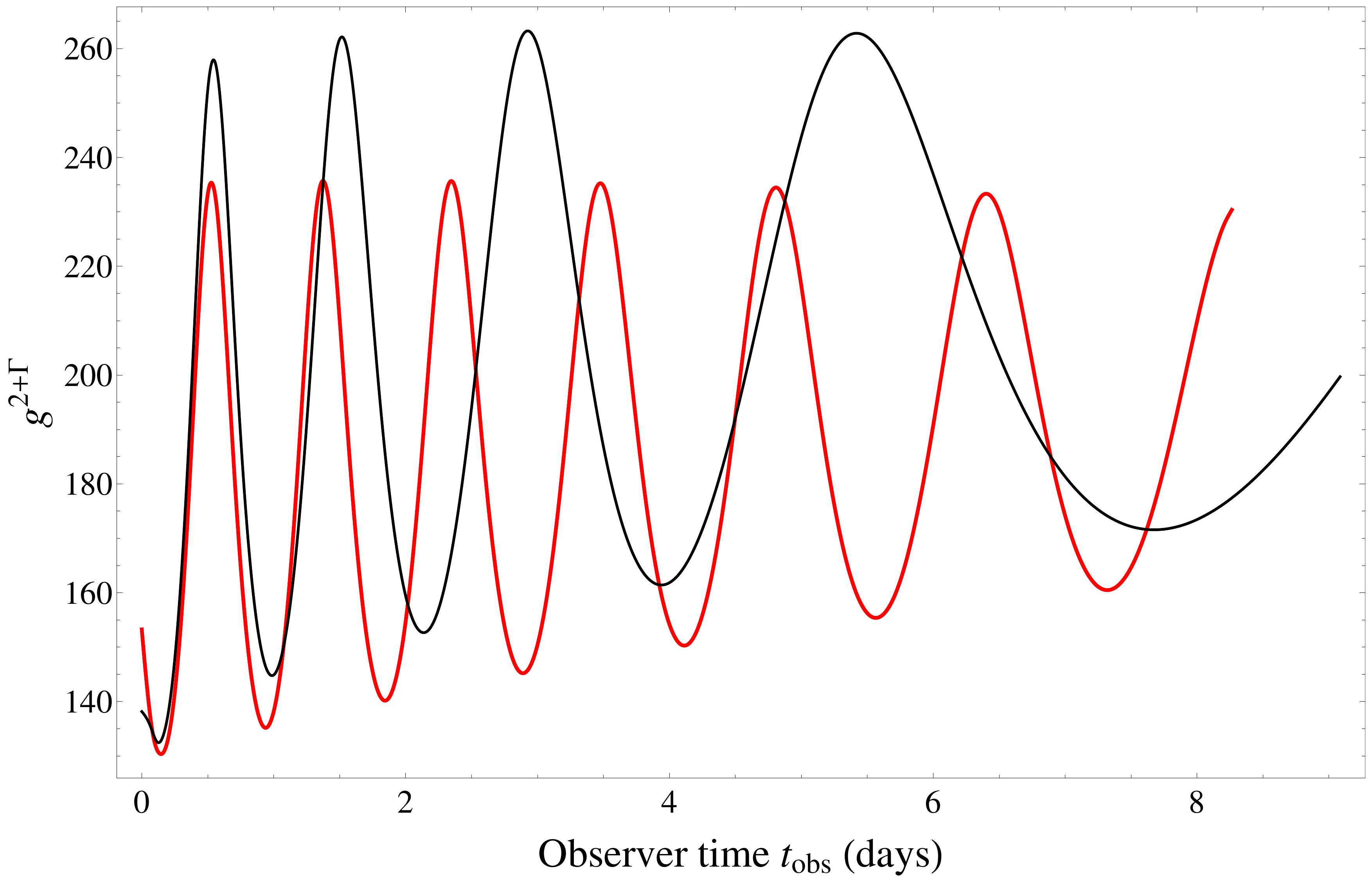}}
\caption{Simulated light curve for $u_p = 3$, $x_A = 0.9$, $i = 0.05^{\circ}$, $M_{\bullet} = 5 \times 10^7 M_{\odot}$, $a = 0.8$, $\varpi_0 = 10 \varpi_L$ and $\Gamma = 1$. The light curve shows the quasi-periodic oscillation expected from a general BL Lac object. A general relativistic light curve from orbital features in a funnel geometry (black curve) is compared with the previously calculated special relativistic light curve from orbital features in a cone geometry (red curve). The increased amplitude ratio and the phase lag change can be attributed to the integrals over time of intensity and phase amplifying the small but early differences in phase. Even a small phase difference between the general relativistic and special relativistic simulations gets amplified due to the nature of the $g$ factor in eqn. (\ref{gfacdef}) which includes the gravitational redshift factor $(1-2 M/R)^{1/2}$ and light bending effects. {\em See the online article for the color version.}}
\label{GRfunnel}
\end{figure}

\begin{figure}[!h]
\centerline{\includegraphics[scale=0.34]{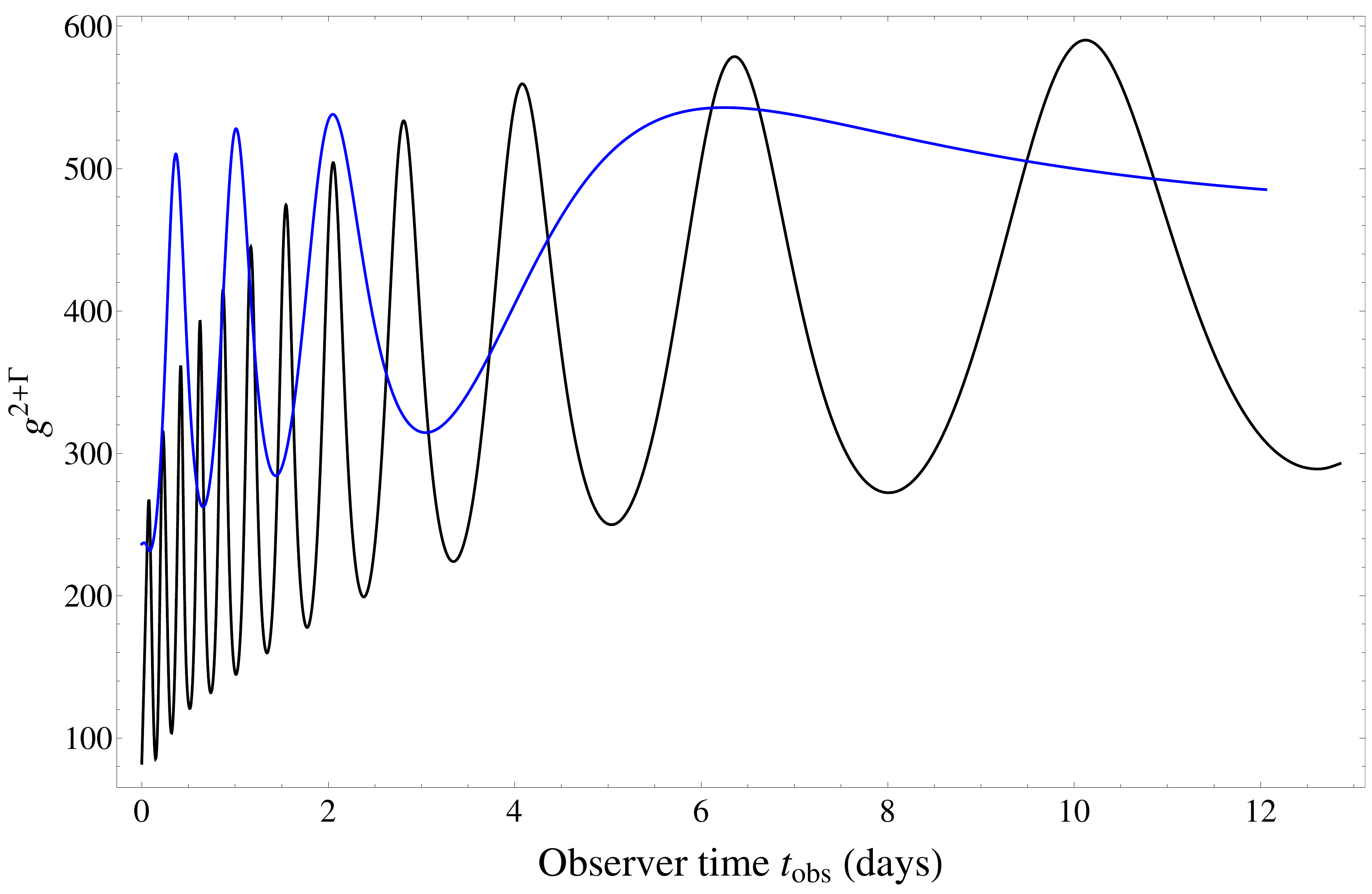}}
\caption{Simulated light curve for $\gamma_{jet,i} = 4$, $k = 2$, $x_A = 0.9$, $i = 0.05^{\circ}$, $M_{\bullet} = 5 \times 10^7 M_{\odot}$, $a$ = 0 and $\varpi_0 = 10 \varpi_L$. A comparison is made between the funnel geometry light curve (black curve) and the cone geometry light curve (blue curve). Both light curves show a quasi-periodic oscillation expected from a general BL Lac object. The phase of the funnel geometry curve systematically lags the phase of the cone geometry curve. Also, the funnel geometry curve is dominated by higher frequencies in the initial portion. {\em See the online article for the color version.}}
\label{GRfc1}
\end{figure}

A comparison is then done between the simulated light curve in funnel and cone geometries for the general relativistic formalism. For the parameter values $\gamma_{jet,i} = 4$, $k = 2$ ($q/f = 3$), $\theta_0 = 0.1^{\circ}$, $i = 5^{\circ}$, $M_{\bullet} = 5 \times 10^7 M_{\odot}$, $a = 0$ and $\varpi_0 = 10 \varpi_L$, the light curves are plotted in Fig. \ref{GRfc1}. The phase of the funnel geometry curve systematically lags the phase of the cone geometry curve indicating a dominance by higher frequencies. This phase shift is seen to increase with the simulation time. The ratio of the maximum amplitude of the funnel geometry curve to that of the maximum amplitude of the cone geometry curve is 9 \%. Because of the slower expansion of the funnel compared to the cone there are more orbits at higher frequencies (due to conservation of angular momentum) at a location where the GR boost factor is more effective.

%At a given time $t$, for the case of $k = 2$, the funnel shape is likely to have already developed into the cylindrical portion. While, at the same time, for the case of a higher $k$, the funnel shape is likely to be still developing. Thus, in the smaller $k$ case, it is expected that the phase shift between the funnel model curve and the cone model curve is larger as the cone structure no longer follows the funnel structure. This is evident in a simulation with the same parameters as above but with $k = 9$ ($q/f = 10$), plotted in Fig. \ref{GRfc2}. We see that the phase difference between the two cases is smaller and is only gradually increasing. Also, the amplitude ratio is very small compared to the previous case. 

\subsection{Funnel model simulations and discussion}
\label{results}

We perform two sets of simulations to address the expected light curve and timing information for an observer viewing the AGN at varying inclination angles $i$, black hole mass $M_{\bullet}$ and the initial launch bulk Lorentz factor $\gamma_{jet,i}$. The choices for $i$ are based on typical values expected for the inclination angle from observational studies (e.g. \citealt{2009A&A...507L..33P}). A time series analysis of the light curve is conducted using the Fourier periodogram analysis (e.g. \citealt{2014JApA...35..397M,2014ApJ...791...74M}) and the wavelet analysis (e.g. \citealt{1998BAMS...79...61T,2011JApA...32..117M}). The Fourier periodogram of the light curve is the normalized Fourier power spectrum, evaluated at the frequencies $f = j/(N \Delta t)$ where $j = 1, 2, ... (N/2-1)$ (upto and excluding the Nyquist frequency) and is fit with a power law shape, $P(f) = A f^{m}$, assumed to be the shape that best describes the underlying power spectral density (PSD). The wavelet analysis gives the QPO and its evolution including the phases during which it is present in the light curve and hence, the number of cycles it is present for. In the wavelet analysis, any periodicity inside a cone of influence (triangular region) can be trusted. Features outside this region could be subject to systematic effects due to the wavelet method. From these simulations, we aim to address questions such as the typical QPO expected from orbital processes in the jet, its dependence on $M_{\bullet}$, the beaming effect, the conditions for its sustenance and the typical range of PSD slopes expected. These are motivated from observational studies at optical and radio wavelengths which indicate QPOs with timescales of less than a day to a few tens of days (e.g. \citealt{2012MNRAS.425.1357G}). 

\begin{figure}[!h]
\centerline{\includegraphics[scale=0.18]{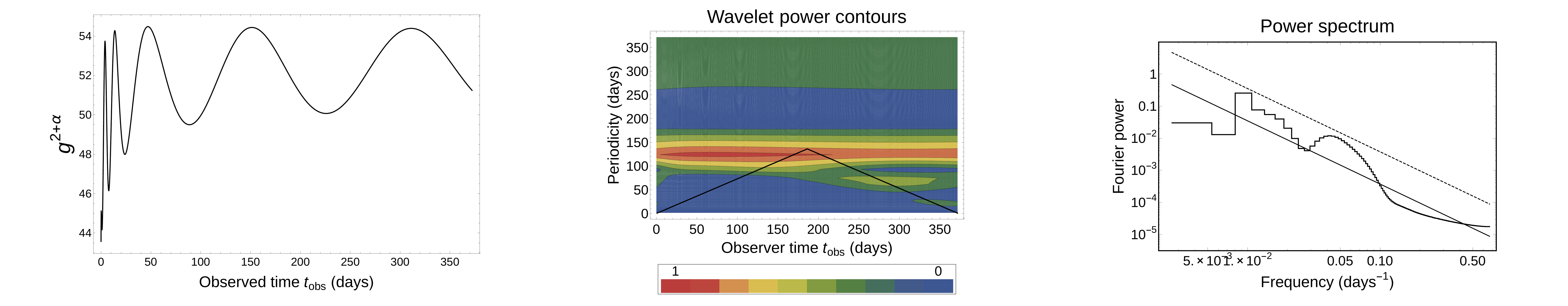}}
\centerline{\includegraphics[scale=0.18]{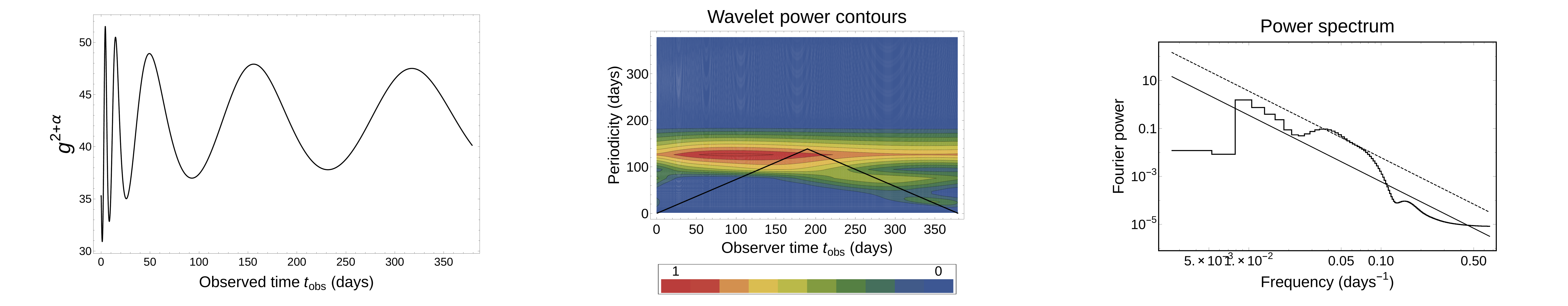}}
\centerline{\includegraphics[scale=0.18]{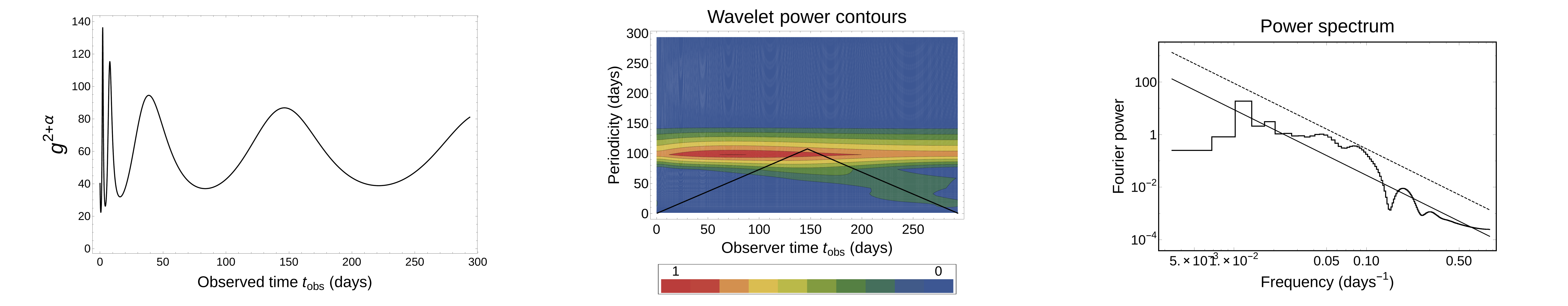}}
\caption{Results for light curves from a single emitting ring of $\varpi_0 = 10 \varpi_L$. The plots in the left column are the simulated light curves, the plots in the middle column are their wavelet analysis and the plots in the right column are their PSD. The PSD is fit with a power law with a slope $m$. The line above the best fit indicates a 99\% significance on any inferred QPO period. Top row: simulation for $i = 3^{\circ}$, $M_{\bullet}$ = $5 \times 10^7 M_{\odot}$, $\gamma_{jet} = 2$; a QPO of 124 days is inferred from the timing analyses; a PSD slope of $-1.96$ is inferred. Middle row: simulation for $i = 9^{\circ}$, $M_{\bullet} = 5 \times 10^7 M_{\odot}$, $\gamma_{jet} = 2$; a QPO of 126.2 days is inferred from the timing analyses; a PSD slope of $-2.78$ is inferred. Bottom row: simulation for $i = 15^{\circ}$, $M_{\bullet} = 5 \times 10^7 M_{\odot}$, $\gamma_{jet} = 4$; a QPO of 97 days is inferred from the timing analyses; a PSD slope of $-2.50$ is inferred. {\em See the online article for the color version.}}
\label{sim1plot}
\end{figure}

We first simulate the expected light curve from an emitting source in helical motion along the funnel shaped magnetic surface launched from a single ring with cylindrical radius $\varpi = 10 \varpi_L$. The other quantities which are fixed include $k = 2$, $x_A = 0.9$, $\theta_0 = 0.1^{\circ}$ and $\alpha = 1$. The simulations are carried out for $i =$ $3^{\circ}$, $6^{\circ}$, $9^{\circ}$, $12^{\circ}$ and $15^{\circ}$ with $M_{\bullet} = (0.5, 5) \times 10^8 M_{\odot}$ and $\gamma_{jet,i} =$ 2 (mildly relativistic), 4 (relativistic) and 10 (highly relativistic). The choice of $\gamma_{jet,i} = 2, 4$ is consistent with the range predicted by the effects of radiation pressure and drag in \S \ref{radprdynamics}. The choice of $\gamma_i = 10$ was made in order to study simulated light curves for jets from black holes with higher masses which includes blazars. The results of the timing analysis are summarized in Table \ref{sim1tab} and some interesting cases are plotted in Fig. \ref{sim1plot}. 

\begin{table}[!h]
\centering
\begin{tabular}{|l|l|l|l|l|l|}
\hline
Inclination & & $\gamma_{jet,i}$ & Max. & PSD & QPO \\ 
Angle &$\disp{\frac{M_{\bullet}}{10^8 M_{\odot}}}$ & & Amplitude & Slope & (days) \\
$i$ ($^{\circ}$)& & & $A$ & $m$ & \\ \hline
3 & 0.5 & 2 & 54.48 & -1.96 & 124.0 W \\
  &     & 4 & 506.49 & -1.97 & 89.8 W \\
  &     & 10 & 8090.84 & -2.26 & 79.7 W \\
  & 5   & 2 & 54.24 & -1.54 & 68.2 \\
  &     & 4 & 492.43 & -1.65 & 18.6 \\
  &     & 10 & 6949.31 & -1.82 &  \\
6 & 0.5 & 2 & 54.05 & -2.42 & 124.8 W \\
  &     & 4 & 483.05 & -2.18 & 90.8 W \\
  &     & 10 & 5780.68 & -3.00 & 80.8 W \\
  & 5   & 2 & 54.06 & -1.64 & 70.1 \\
  &     & 4 & 483.10 & -1.73 & 20.7 \\
  &     & 10 & 5780.68 & -2.01 &  \\
9 & 0.5 & 2 & 51.50 & -1.70 & 126.2 W \\
  &     & 4 & 372.24 & -2.30 & 92.4 W \\
  &     & 10 & 1540.67 & -3.16 & 82.5 W \\
  & 5   & 2 & 51.51 & -1.70 & 73.2 \\
  &     & 4 & 372.25 & -1.77 & 24.3 \\
  &     & 10 & 1544.25 & -2.03 &  \\
12 & 0.5 & 2 & 46.60 & -1.73 & 128.2 W\\
  &     & 4 & 238.24 & -2.42 & 94.8 \\
  &     & 10 & 328.10 & -3.13 & 85.0 W \\
  & 5   & 2 & 46.60 & -1.73 & 77.5  \\
  &     & 4 & 238.24 & -1.80 & 29.0 \\
  &     & 10 & 328.56 & -1.98 & - \\
15 & 0.5 & 2 & 40.24 & -3.43 & 130.7 W \\
  &     & 4 & 136.38 & -2.50 & 97 W \\
  &     & 10 & 81.36 & -3.07 & 88.1 W \\
  & 5   & 2 & 40.25 & -1.75 & 82.9 \\
  &     & 4 & 136.38 & -1.82 & 35.1 \\
  &     & 10 & 81.44 & -1.94 &  \\ \hline 
\end{tabular}
\caption{Results for light curves from blobs launched from a single ring of cylindrical radius $\varpi_0 = 10 \varpi_L$. The parameters used which are not reported in the above table include $x_A = 0.9$, $\theta_0 = 0.1^{\circ}$ and $\alpha = 1$. The slopes steeper than $-2.5$ are likely to be the result of a poor fit. Hence, we calculate a median slope of $-2.00 \pm 0.28$ and the QPO estimates range between 20.7 - 130.7 days. A $W$ next to a QPO peak indicates its measurement using the wavelet analysis only. If there is nothing present next to the inferred QPO, the measurement is the mean of the wavelet and PSD estimates.}
\label{sim1tab}
\end{table}

With increase in the inclination angle $i$, the maximum amplitude in the simulated light curve tends to decrease. This is due to the beaming effect as the projection of the emitting source velocity onto the emission direction (${\bf \beta_{jet}\cdot k}$) continues to grow smaller in the direction towards the observer. This leads to a less pronounced maximum amplitude which continues to decrease with increase in $i$. For a given $M_{\bullet}$, the maximum amplitude tends to increase with $\gamma_{jet,i}$. This is due to the stronger beaming effect towards the observer line of sight by the emitting source for larger $\gamma_{jet,i}$. The PSD slope ranges between $-1.54$ and $-3.43$. The slopes steeper than $-2.5$ are likely to be the result of a poor fit. Hence, we calculate a median slope of $-2.00 \pm 0.28$. QPOs peaked between 20.7 - 130.7 days are inferred from the wavelet and PSD analysis. For a given $M_{\bullet}$, the QPO peak shifts to higher values with increasing $i$. This is due to the orientation away from the observer's line of sight for the chosen set of parameters leading to the larger timescale QPO at larger $r$.  

\begin{figure}[!h]
\centerline{\includegraphics[scale=0.18]{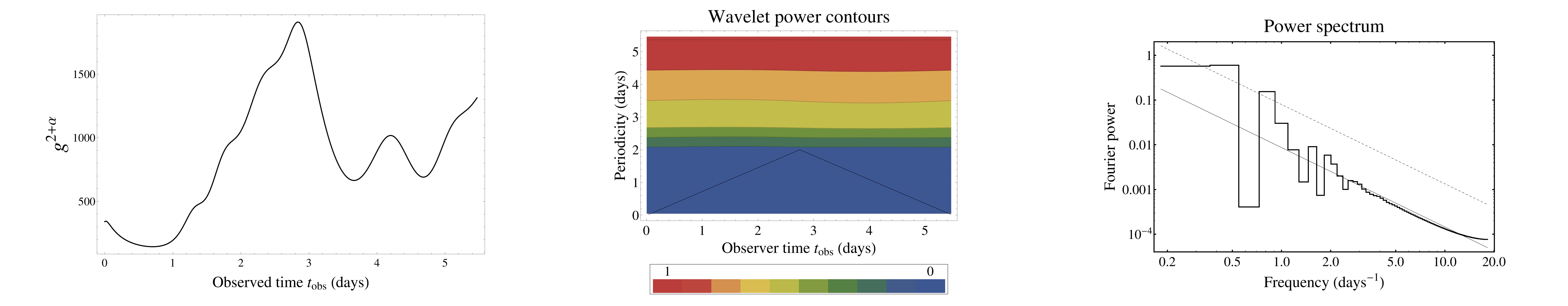}}
\centerline{\includegraphics[scale=0.18]{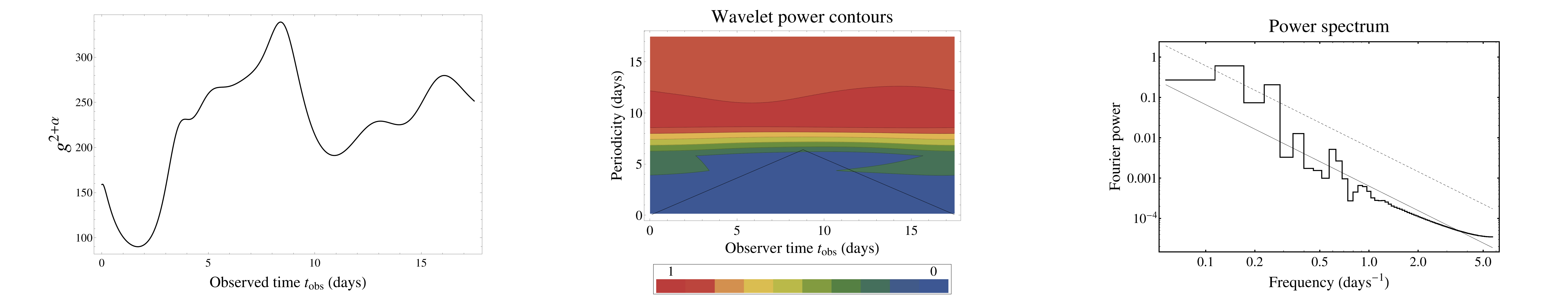}}
\centerline{\includegraphics[scale=0.18]{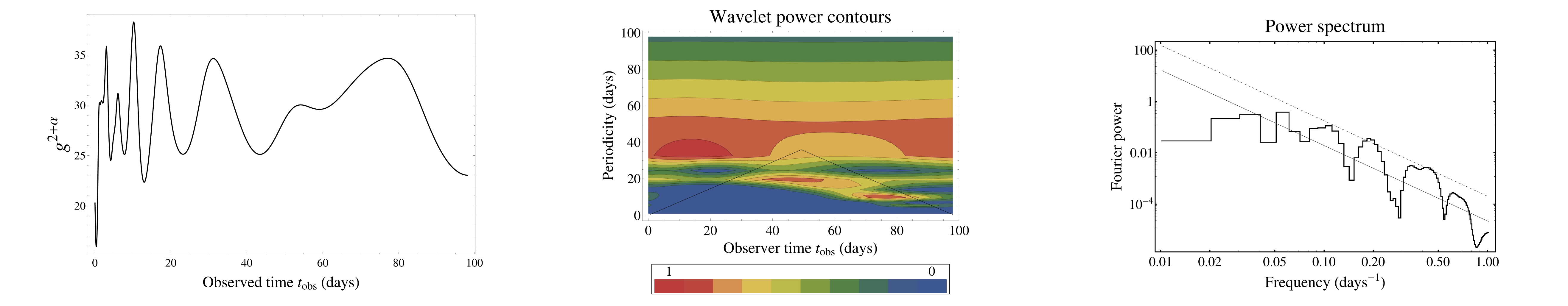}}
\caption{Results for light curves from multiple emitting rings of $\varpi_0 =$ $5 \varpi_L$ to $8 \varpi_L$. The plots in the left column are the simulated combined light curves. The plots in the middle column are their wavelet analysis and the plots in the right column are their PSD. The PSD is fit with a power law with a slope $m$. Top row: simulation for $i = 3^{\circ}$, $M_{\bullet}$ = $5 \times 10^8 M_{\odot}$, $\gamma_{jet} = 10$; a QPO of 1.37 days and a PSD slope of $-1.77$ are inferred from the timing analyses. Middle row: simulation for $i = 6^{\circ}$, $M_{\bullet} = 5 \times 10^8 M_{\odot}$, $\gamma_{jet} = 4$; QPOs of 4.37 days and 8.74 days (harmonics) and a PSD slope of $-2.12$ are inferred from the timing analyses. Bottom row: simulation for $i = 15^{\circ}$, $M_{\bullet} = 5 \times 10^7 M_{\odot}$, $\gamma_{jet} = 2$; a QPO of 5.45 days and a PSD slope of $-2.94$ are inferred from the timing analyses. {\em See the online article for the color version.}}
\label{sim2plot}
\end{figure}

We then simulate the expected light curve from multiple emitting sources in helical motion along the funnel shaped magnetic surface launched from rings with cylindrical radii $\varpi$ = 5, 6, 7 and 8 $\varpi_L$. Fixed parameters in this simulation include $k = 2$, $x_A = 0.9$, $\theta_0 = 0.1^{\circ}$ and $\alpha = 1$. The simulations are carried out for $i$ = $3^{\circ}$, $6^{\circ}$, $9^{\circ}$, $12^{\circ}$ and $15^{\circ}$ with $M_{\bullet} = (0.5, 5) \times 10^8 M_{\odot}$ and $\gamma_{jet,i} = 2 - 10$ as in the previous case. The results of the timing analysis are summarized in Table \ref{sim2tab} and some interesting cases are plotted in Fig. \ref{sim2plot}.

\begin{table}[!h]
\centering
\begin{tabular}{|l|l|l|l|l|l|}
\hline
Inclination & & $\gamma_{jet,i}$ & Max. & PSD & QPO  \\ 
Angle & $\disp{\frac{M_{\bullet}}{10^8 M_{\odot}}}$ &  & Amplitude & Slope & (days) \\
$i$ ($^{\circ}$)& & & $A$ & $m$ & \\ \hline
3 & 0.5 & 2 & 54.35 & -2.01 & - \\
  &     & 4 & 494.97 & -1.67 & - \\
  &     & 10 & 5955.14 & -1.70 & - \\
  & 5   & 2 & 47.94 & -1.72 & 27.62 \\
  &     & 4 & 327.01 & -1.90 & 7.89 \\
  &     & 10 & 1909.06 & -1.77 & 1.37 \\
6 & 0.5 & 2 & 52.28 & -2.47 & - \\
  &     & 4 & 424.37 & -1.73 & - \\
  &     & 10 & 4155.13 & -1.96 & 0.98 \\
  & 5   & 2 & 47.43 & -1.95 & 28.34\\
  &     & 4 & 338.99 & -2.02 & 4.37, 8.74 \\
  &     & 10 & 3553.40 & -2.36 & 1.81 \\
9 & 0.5 & 2 & 48.46 & -2.81 & - \\
  &     & 4 & 322.86 & -1.81 & - \\
  &     & 10 & 2778.43 & -1.31 & - \\
  & 5   & 2 & 45.01 & -2.12 & 14.80, 29.59 \\
  &     & 4 & 298.02 & -2.18 & 5.06, 10.13 \\
  &     & 10 & 2778.43 & -3.18 & - \\
12 & 0.5 & 2 & 44.00 & -2.99 & - \\
  &     & 4 & 217.66 & -2.00 & - \\
  &     & 10 & 962.85 & -1.09 & - \\
  & 5   & 2 & 40.96 & -2.39 & 15.65, 31.31 \\
  &     & 4 & 218.57 & -2.34 & 6.03 \\
  &     & 10 & 962.85 & -3.39 & - \\
15 & 0.5 & 2 & 38.23 & -2.94 & 5.45 \\
  &     & 4 & 135.59 & -1.97 & - \\
  &     & 10 & 207.60 & -1.23 & - \\
  & 5   & 2 & 35.78 & -2.61 & 16.74, 33.49 \\
  &     & 4 & 137.58 & -2.46 & 2.89, 7.22 \\
  &     & 10 & 207.60 & -3.24 & 1.89 \\ \hline 
 \hline 
\end{tabular}
\caption{Results for light curves from blobs launched from multiple rings of cylindrical radii $\varpi_0 = 5-8 \varpi_L$. The parameters used which are not reported in the above table include $x_A = 0.9$, $\theta_0 = 0.1^{\circ}$ and $\alpha = 1$. We calculate a median slope of $-2.02 \pm 0.34$ and the QPO estimates range between 1.37 - 33.49 days. All QPOs inferred above are the mean of the wavelet and PSD estimates.}
\label{sim2tab}
\end{table}

In these simulations too, we observe similar trends as were present in the previous case. With increase in $i$, the maximum amplitude in the simulated light curve tends to decrease. For a given $M_{\bullet}$, the maximum amplitude tends to increase with $\gamma_{jet,i}$. The PSD slope in these simulations is more well constrained and ranges between $-1.09$ and $-3.39$. The slopes steeper than $-2.5$ are likely to be the result of a poor fit. Hence, we calculate a median slope of $-2.02 \pm 0.34$. Significant QPOs ranging between 1.37 - 33.49 days are detected in these simulations. The QPO for high $\gamma_{jet,i}$ tends to be lower than that for lower $\gamma_{jet,i}$. This is likely as for high $\gamma_{jet}$, the beamed portion occurs in the inner jet close to the black hole where the orbital frequency is higher due to gravitational redshift. Thus, the associated timescales are then expected to be lower. Thus, the power spectrum is dominated by higher frequencies which implies that the PSD would tend to flatten. This trend is also seen in the above multiple emitting ring simulations for $M_{\bullet}$ = $5 \times 10^7 M_{\odot}$. For the case of $M_{\bullet} = 5 \times 10^8 M_{\odot}$, the opposite trend is seen. For a higher $M_{\bullet}$, there are more cycles during beaming due to favourable orientation (at higher frequencies); hence, the PSD slope tends to steepen.

\section{Summary \& Conclusions}
\label{conclusions}

We summarize our results. The special relativistic cone model (\S \ref{SRcone}) is the same as that proposed by \cite{1992A&A...255...59C}. The fully relativistic models developed in this work include the cone model (\S \ref{GRcone}) and the funnel model (\S \ref{GRfunnel}). The main results include discussions on the novel aspects of the fully relativistic funnel model with respect to the special relativistic cone model.

\begin{enumerate}

\item 
We have considered a possible mechanism of jet based variability from various sources including BL Lac objects and blazars (e.g. \citealt{1992A&A...255...59C,1995A&A...302..335S}), quasars (e.g. \citealt{2011A&A...526A..51K}), binary black holes (e.g. \citealt{2010ApJ...724L.166I}) and X-ray binaries (e.g. \citealt{2008ApJ...679.1413F,2009ApJ...695.1199F}). We extended the light-house model proposed in \cite{1992A&A...255...59C} which was applied to simulate optical light curves from BL Lac objects and quasars. 

\item 
We studied the effects of radiation pressure and drag and derived the saturation Lorentz factors that are achievable. The Lorentz factors $\gamma$ range between $2 - 7$ in the purely radial outflow simulations. For a poloidal flow ($\beta_\phi = 0$), we obtain $\gamma$ in the range $1.1 - 26.3$, greater than that obtained previously. Though, the drag force acting on the non-radial component of the blob velocity plays an important role in rapidly decreasing $\gamma \sim 2$ thus stabilizing it at very small $x$, indicating that the outflow is stable even in the innermost regions. Thus, the $\gamma_{jet,i}$ chosen for the simulations in the relativistic cone and funnel models in the subsequent section can be obtained at larger distances with small to moderately relativistic emitting blob velocities.

\item
We constructed a kinematic model of orbital blobs along helical trajectories on a magnetic surface approximating the expanding jet with foot points on the accretion disk in the vicinity of a black hole using a special relativistic calculation of the $g$ factor for cone geometry similar to \citep{1992A&A...255...59C} in \S \ref{SRcone} and a fully relativistic formulation in cone geometry (\S \ref{GRcone}) and the same in a more realistic funnel geometry (\S \ref{GRfunnel}). The $g$ factor was calculated in Schwarzschild geometry and the following GR effects were included: gravitational and Doppler shifts, aberration and a prescription for time delay due to orientation and light bending.

\item 
By using the periodogram and wavelet based time series analysis techniques (\S \ref{results}), we seek to distinguish amongst the various flavours of the generic jet variability model which here include the cone and funnel models. The time series analysis of simulated light curves yields properties including aperiodic variability, its timescales and the emission region; the power spectral density slope and its range which can be compared with observations in optical/UV and X-ray; it is also found that the QPO, its evolution and stability can be described by the orbital motion of blobs in the jet.

\item
The use of fully relativistic models (given in \S \ref{GRcone} and \S \ref{GRfunnel}) was justified by the resulting amplitude increase (by about 12 \%) due to the GR boosts and a systematically increasing phase lag in the simulated light curve for a general BL Lac object from the fully relativistic funnel when compared to the special relativistic cone which occurs at small $k$ and larger inclination angles $i$. The phase lag has been explained by the angular momentum conservation in combination with the gradual increase of the orbital radius. Our fully relativistic formulation reduces to the special relativistic formulation in the limit of large $R$ and when all the above GR effects are not considered; this is presented in \S \ref{kinmodels}. 

\item
There is thus a necessity to account for all GR effects as used in the present model for the correct description of the physical effects on emitted radiation from these jetted sources and to use more realistic jet geometries. It can be thus applied to both timing studies of jet variability as well as to map trajectories of radio blobs or blobs in the inner jet and compared with observations to help identify the region of emission. The application of the model to radio data is being planned.

\item
Two sets of simulations, one for blobs emanating from a single and another from multiple rings were carried out to span a range of light curves. Their timing properties were studied for the maximum amplitudes, PSD slope and QPO if present. Two main trends are observed in the single emitting ring simulations. With increase in the inclination angle $i$, the maximum amplitude in the simulated light curve tends to decrease, as the beaming effect in the direction of the observer continues to grow smaller. The other trend is that for a given $M_{\bullet}$, the maximum amplitude tends to increase with $\gamma_{jet,i}$, due to the stronger beaming effect towards the observer line of sight by the emitting source for larger $\gamma_{jet,i}$. We calculate a median slope of $-2.00 \pm 0.28$ similar to the observed PSD slopes of blazars in the Optical and X-ray translated to the source frame. QPOs peaked between  20.7 - 130.7 days are inferred from the wavelet and PSD analysis. The second set of simulations were carried out for an blob launched from multiple emitting rings $\varpi_0 = 5 - 8 \varpi_L$. In these multiple emitting rings simulations too, the same above two trends are observed. We calculate a median slope of $-2.02 \pm 0.34$. Significant QPOs ranging between 1.37 - 33.49 days are detected in these simulations. In the multiple emitting rings simulations, as we were also able to identify QPOs as a wider choice of orientations result from a larger set of initial conditions; same trends were observed in the relationship between the QPO timescale, $M_{\bullet}$ and $\gamma_{jet}$. The QPO timescale tends to reduce with increase in $\gamma_{jet}$ due to GR boost which tends to flatten the PSD slope. Aperiodic variability when considering multiple emitting rings is a natural consequence that is seen in our results.

%have been missed in the special relativistic cone model of \cite{1992A&A...255...59C}.}

\end{enumerate}

A natural power law shaped PSD with a typical slope of $\sim$ $-2$ along with a weak to strong QPO ranging between 1.37 - 130.7 days emerges from the simulations considering single and multiple emitting rings. This shape comes naturally considering only geometrical parameters, even before any consideration of physical models of the orbital instabilities. The detailed magnetic structure can be constructed by solving the relativistic Grad-Shafranov equation. The dynamical timescale due to jet based orbital processes is expected to be from a few minutes to tens of days depending on the black hole mass. The general relativistic funnel model for variability can thus be applied to radio, optical and X-ray emission from various types of jetted sources including radio loud AGN such as BL Lac objects, quasars, binary black holes, X-ray binaries where active accretion results in the jet phenomenon such as in micro-quasars and other compact sources such as accreting and hence active neutron stars.

\section*{Acknowledgements}

We are indebted to the anonymous referee for  suggesting the incorporation of effects of radiation pressure and drag and other improvements that have enhanced this paper considerably. We thank V. K. Subramanian for preparing schematic Figs. \ref{RadiationGeometry}, \ref{emmgeom} and \ref{funnelgeom}.

\newpage
\begin{appendix}

\section{Relationship between the effective redshift factor $g$ and the Doppler factor $D$}
\label{gfacderivation}

The tetrads used by a local static observer in the Schwarzschild metric are given by,

\begin{align}\label{tetrads}
e^{\mu}_{(t)} &= (1-2 M/R)^{-1/2} (1,0,0,0) \\ \nonumber
e^{\mu}_{(r)} &= (1-2 M/R)^{1/2} (0,1,0,0) \\ \nonumber
e^{\mu}_{(\theta)} &= (1/R) (0,0,1,0) \\ \nonumber
e^{\mu}_{(\phi)} &= (1/R \sin \theta) (0,0,0,1).
\end{align}
%The tetrads used by a local static observer in the Schwarzschild metric to measure physical quantities are presented in Appendix \ref{refframes}. 
The directional vector $n^{(a)}$ in the general relativistic case with light bending is given by \citep{2005A&A...441..855P},
\begin{equation}
n^{(a)} = (n^{(t)},n^{(r)},n^{(\theta)},n^{(\phi)}) = \left(1,\cos \alpha,- \frac{\sin \alpha}{\sin \psi} \cos i,- \frac{\sin \alpha}{\sin \psi} \sin \phi \sin i \right).
\label{emissionvec}
\end{equation}

If ${\bf p}$ is the four-momentum of the emitted light ray, its covariant components are given 
\begin{equation}
p_{\mu} = p_t (1,p_r/p_t,p_\theta/p_t,p_\phi/p_t),
\label{lrpmu}
\end{equation} 
the objective being to calculate the components $p_{\mu}$ in terms of the emission angles and the geometric quantities. This can be done using the relationship between the directional vector $n^{(a)}$ and $p_{\mu}$ given by
\begin{equation}
n^{(a)} = \frac{p_\mu e^{\mu}_{(a)}}{p_\mu e^{t}_{(a)}}.
\end{equation}

Using the above equation and the tetrads from eqn. (\ref{tetrads}), we obtain
\begin{align}\label{lr4mom}
p_r/p_t &= \cos \alpha (1-2 M/R)^{-1}\\ \nonumber
p_\theta/p_t &= - \frac{\sin \alpha}{\sin \psi} \cos i R (1-2 M/R)^{-1/2} \\ \nonumber
p_\phi/p_t &= - \frac{\sin \alpha}{\sin \psi} \sin \phi \sin i R \sin \theta (1-2 M/R)^{-1/2} 
\end{align}

The four velocity of the emitting source is given by
\begin{equation}
{\bf u} = u^t (1, \beta_r,\beta_{\theta},\beta_{\phi}),
\label{sourceu}
\end{equation}
where $u^t = \gamma_{jet}$ is obtained from eqn. (\ref{gammajet}). The effective redshift factor $g$ is the ratio of observed to emitted energy of the photon and is given by,
\begin{equation}
g = \frac{E_{\mathrm{observed}}}{E_{\mathrm{emitted}}} = \frac{{\bf p}(\infty)\cdot {\bf u}(\infty)}{{\bf p}\cdot {\bf u}} = \frac{-1}{p_{\mu} u^{\mu}}.
\label{gfacdef1}
\end{equation}

Using eqns. (\ref{lr4mom}) and (\ref{sourceu}) in the above equation, the effective redshift factor is given by, 

\begin{align} 
g &= \frac{(1-2 M/R)^{1/2}}{u^t} \left((1-2 M/R)^{1/2}-\cos \alpha \frac{\beta_r}{(1-2 M/R)^{1/2}}\right. \\ \nonumber
&\left.-\frac{\sin \alpha}{\sin \psi} R \beta_\theta \cos i-\frac{\sin \alpha}{\sin \psi} R \sin \theta \beta_\phi \sin \phi \sin i\right)^{-1}.
\label{gfactorut}
\end{align}

The $g$ expression can be written in terms of $\beta_{jet} \cos \xi$ with $\beta_{jet}$ obtained from eqn. (\ref{betajet}) already including a correction for the frame changing from an instantaneous rest frame to a local static observer frame and $\cos \xi$ obtained from eqn. (\ref{cosxi}) to give the expression for $g$ in eqn. (\ref{gfacdef}). Thus,
\begin{equation}
g = \frac{(1-2 M/R)^{1/2}}{\gamma_{jet} (1-\beta_{jet} \cos \xi)}.
\end{equation}

As $p_\mu$ can also be expressed as 
\begin{equation}
p_\mu = \epsilon (1,b (1/b^2-1/R^2 (1-2 M/R))^{-1/2},p_{\theta}/\epsilon,b),
\end{equation}
where $\epsilon$ is the conserved photon energy, the direction between the emitted ray and the radial direction for equatorial orbits ($\theta = \pi/2$) is given by
\begin{equation}
\tan \alpha = \frac{p_\mu e^{\mu}_{(\phi)}}{p_\mu e^{\mu}_{(r)}} = \frac{(b/R)}{(1-2 M/R))^{-1/2} b (1/b^2-1/R^2 (1-2 M/R))^{-1/2}}.
\end{equation}
Solving the above equation for $\sin \alpha$, we obtain eqn. (\ref{sinalpha}).

\end{appendix}

\bibliography{main}

\end{document}